\documentclass{aa}
\usepackage{psfig}
\usepackage{txfonts}
\usepackage{natbib}
\usepackage{amssymb}
\bibpunct{(}{)}{;}{a}{}{,}

\def\pcsk{ph\,cm$^{-2}$\,s$^{-1}$\,keV$^{-1}$}
\def\ecs{erg\,cm$^{-2}$\,s$^{-1}$}
\def\es{erg\,s$^{-1}$}

\def\chir{$\chi^2_r$}

\def\igr{INTEGRAL}
\def\xmm{XMM-{\it Newton}}

\def\pca{RXTE-PCA}
\def\asca{ASCA-GIS}

\def\0142{4U~0142+61}
\def\axp1048{1E~1048.1-5937}

\def\rxs1708{1RXS~J1708-40}
\def\xte1810{XTE~J1810-197}
\def\1841{1E~1841-045}
\def\ax1845{AX~J1845.0-0258}
\def\2259{1E~2259+586}

\begin{document}
\bibliographystyle{aa}

\title{Detailed high-energy characteristics of AXP 4U~0142+61}

\subtitle{Multi-year observations with \igr, RXTE, \xmm\ and ASCA}
\titlerunning{Detailed high-energy characteristics of AXP 4U~0142+61}

\author{P.R. den Hartog\inst{1}
  \and L. Kuiper\inst{1}
  \and W. Hermsen\inst{1,2}
  \and V.M. Kaspi\inst{3}
  \and R. Dib\inst{3}
  \and J. Kn\"odlseder\inst{4}
  \and F.P. Gavriil\inst{5}}

\offprints{P.R. den Hartog}
\mail{Hartog@sron.nl}

\institute{SRON, Netherlands Institute for Space Research,
  Sorbonnelaan 2, 3584 CA Utrecht, The Netherlands
  \and Sterrenkundig Instituut Anton Pannekoek, University of Amsterdam,
  Kruislaan 403, 1098 SJ Amsterdam, The Netherlands
  \and Physics Department, McGill University,
  3600 University Street, Montreal, PQ H3A 2T8, Canada
  \and Centre d'\'Etude Spatiale des Rayonnement, CNRS/UPS, 31028
  Toulouse Cedex 4, France
  \and NASA Goddard Space Flight Center, Astrophysics Science Division,
  Code 662, X-ray Astrophysics Laboratory, Greenbelt, MD 20771, USA}

\date{Received 14 January 2008 / Accepted 3 April 2008}

\abstract{\0142 is one of the Anomalous X-ray Pulsars exhibiting hard
  X-ray emission above 10~keV discovered with \igr.  In this paper we
  present detailed spectral and temporal characteristics both in the
  hard X-ray ($>$10~keV) and soft X-ray ($<$10~keV) domains, obtained
  using data from \igr, \xmm, ASCA and RXTE. Accumulating data
  collected over four years with the imager IBIS-ISGRI aboard \igr,
  the time-averaged total spectrum shows a power-law like shape with
  photon index $\Gamma = 0.93 \pm 0.06.$ \0142 is detected up to
  229~keV and the flux between 20~keV and 229~keV is $(15.01 \pm 0.82)
  \times 10^{-11}$ \ecs, which exceeds the energy flux in the
  2--10~keV band by a factor of $\sim$2.3. Using simultaneously
  collected data with the spectrometer SPI of \igr\, the combined
  total spectrum yields the first evidence for a spectral break above
  100 keV: Assuming for the spectral shape above 20~keV a logparabolic
  function the peak energy of \0142 is $228^{+65}_{-41}$~keV. There is
  no evidence for significant long-term time variability of the total
  emission from \0142. Both the total flux and the spectral index are
  stable within the 17\% level (1$\sigma$).  Pulsed emission is
  measured with ISGRI up to 160~keV. The 20--160~keV profile shows a
  broad double-peaked pulse with a 6.2$\sigma$ detection significance.
  The total pulsed spectrum can be described with a very hard
  power-law shape with a photon index $\Gamma = 0.40 \pm 0.15$ and a
  20--150 keV flux of $(2.68 \pm 1.34) \times 10^{-11}$ \ecs. To
  perform accurate phase-resolved spectroscopy over the total X-ray
  window, we produced pulse profiles in absolute phase for \igr-ISGRI,
  \pca, \xmm-PN and \asca.  The two known pulses in all soft X-ray
  profiles below 10~keV are located at the same phases. Three
  \xmm\ observations in 2003--2004 show statistically
  identical profiles. However, we find a significant profile
  morphology change between an \asca\ observation in 1999 following a
  possible glitch of \0142. This change can be accounted to
  differences in relative strengths and spectral shapes (0.8--10 keV)
  of the two pulses. The principle peak in the \igr\ pulse profile
  above 20~keV is located at the same phase as one of the pulses
  detected below 10~keV.  The second pulse detected with \igr\ is
  slightly shifted with respect to the second peak observed in the
  soft X-ray band. We performed consistent phase-resolved spectroscopy
  over the total high-energy band and identify at least three
  genuinely different pulse components with different spectra.  The
  high level of consistency between the detailed results from the four
  missions is indicative for a remarkable stable geometry underlying
  the emission scenario.  Finally, we discuss the derived detailed
  characteristics of the high-energy emission of \0142 in relation to
  three models for the non-thermal hard X-ray emission.

}

\maketitle
\section{Introduction}
\label{sec:intro}

Anomalous X-ray Pulsars (AXPs) are a special subset of young isolated
neutron stars with pulse periods in the range $\sim$2--12\,s and with
high period derivatives on the order of
$10^{-10}-10^{-12}$\,s\,s$^{-1}$. Their (anomalous) X-ray luminosities
($L_{\rm X} \sim 10^{33} - 10^{35}$\, \es\, within 2--10 keV) are
orders of magnitude too high to be explained by rotational energy
released due to spin down. Different origins of energy budgets
proposed to explain the high luminosities were not consistent with all
observational characteristics of AXPs \citep[c.f.][ for a review and
  latest developments]{Woods06_review, Kaspi07_london}.  However, for
Soft Gamma-ray Repeaters (SGRs), which are another subset of young
isolated X-ray pulsars with similar periods, a toroidal magnetic field
in the neutron-star core as energy reservoir was proposed by
\citet{DT92}. Their magnetar model for SGRs can explain the anomalous
X-ray luminosities of AXPs by the decay of an ultra-high magnetic
field ($B \gtrsim 10^{15}$G). The magnetar model also explains
numerous of other observable characteristics from AXPs like the large
period derivatives and bursting behaviour. Most features of the
magnetar model can be found in \citet[][]{DT92,TD93,TD95,TD96,TLK02}.
Observational evidence collected over the last five years clearly
showed that AXPs and SGRs have multiply similar characteristics and
therefore are now believed to be subsets of magnetars
\citep{Woods06_review}.

In this paper we will present new results on AXP~\0142. This
particular AXP has a spin period of $P = 8.69$ s and a period
derivative of $\dot{P} = 0.20 \times 10^{-11}$~s\,s$^{-1}$. The
inferred surface magnetic field strength is $B = 3.2 \times 10^{19}\,
\sqrt{P\dot{P}}$~G = $1.3 \times 10^{14}$ G. In the 2--10 keV band this is
the brightest persistent AXP with a luminosity $L_X \sim 1 \times
10^{35}$ \es\, assuming a distance of 3.6 kpc
\citep{Durant06_distances}.

A remarkable discovery with the hard X-ray/soft gamma-ray telescope
INTEGRAL \citep{Winkler03_igr} is the existence of a hard spectral
component in the spectra of AXPs at energies above 20~keV. In 2004
\citet{Molkov04_sagarm} reported hard X-ray emission ($>$20 keV) from
the position of AXP \1841 in the supernova remnant G27.4+0.0
(Kes~73). \citet{Kuiper04_1841} showed unambiguously that the hard
X-rays indeed originate from the AXP and not from the supernova
remnant by detecting pulsations in the hard X-ray emission from this
source using {\em Rossi X-ray Timing Explorer} (RXTE) PCA
and HEXTE data.

\begin{table*}[!t]
\centering
\renewcommand{\tabcolsep}{1.7mm}
\caption[]{Summary of the \igr\, observations of \0142. The following
  information is given: the revolution intervals; the observations
  time spans both in MJD as in calendar dates; the number of Science
  Windows; exposure times ($t_{\rm{exp}}$); and the effective
  on-source exposure ($t_{\rm{eff}}$).}

\begin{tabular}{l r@{ -- }l l@{ }r l@{ -- } l@{ }l l r r@{.}l  r@{.}l}

\vspace{-3mm}\\

\hline
\hline
\vspace{-3mm}\\
Rev. &
\multicolumn{8}{c}{Time span} &
ScWs &
\multicolumn{2}{r}{$t_{\mathrm{exp}}\,(\rm{ks})$} &
\multicolumn{2}{r}{$t_{\mathrm{eff}}\,(\rm{ks})$}\\
\hline
\vspace{-3mm}\\
142--468 & 52985 & 53960 & Dec.&12, &2003 & Aug. &13, &2006 &
 1617 & 4494&4 &2365&3\\
\hline
\vspace{-3mm}\\

142--189 & 52985 & 53125 & Dec.&12, &2003 & Apr. &30, &2004 &
  404 & 963&2 & 342&0\\
202--269 &53165 & 53367  & Jun.&9,  &2004 & 
  Dec. &28, &2004 & 410 & 976&5 & 501&6\\
331--336 & 53550 & 53568 & Jun.&29, &2005 & Jul. &17, &2005 &
  265 & 867&6 & 827&9\\
384--396 & 53710 & 53745 & Dec.&6,  &2005 & Jan. &10, &2006 &
  435 & 1334&3 & 420&6\\
452--468 & 53912 & 53960 & Jun.&26,  &2006 & Aug. &13, &2006 &  
  103 & 353&8 & 273&2\\
\hline
\hline
\vspace{-3mm}\\
454 ToO  & 53918 & 53920 & Jul.&2,  &2006 & Jul. & 4,  &2006 & 
  59 & 213&7 & 209&7\\
528 ToO  & 54139 & 54142 & Feb.&8,  &2007 & Feb. & 11,  &2007 & 
  54 & 198&6 & 195&5\\
\hline
\end{tabular}

\label{tab:obs}
\end{table*}

For two other AXPs INTEGRAL detections have been
reported. \citet{Revnivtsev04_gc} and \citet{denHartog04_atel0142}
detected 1RXS~J170849-400910 (hereafter \rxs1708) and \0142,
respectively. Also for these and for a fourth AXP, \2259,
\citet{Kuiper06_axps} discovered pulsed hard X-ray emission using RXTE
PCA and HEXTE data.

The INTEGRAL total-emission spectra available for \1841, \rxs1708 and
\0142 are all power-law like and hard. The spectra of these three AXPs
have photon indices\footnote{Defining the power law as $F = F_0 \times
  E^{-\Gamma}$ with $F_0$ the normalisation in units of \pcsk and
  $\Gamma$ the photon index} of $1.32 \pm 0.11$, $1.44 \pm 0.45$ and
$1.05 \pm 0.11$, respectively \citep{Kuiper06_axps}. Furthermore,
\citet{denHartog07_london} performed a multi-frequency campaign in
which \0142 was observed from radio to hard X-rays almost
simultaneously. A dedicated INTEGRAL observation of 1~Ms was part of
this campaign. This observation yielded a power-law like spectrum up
to 229~keV with a photon index of $0.79 \pm 0.10$ with a luminosity of
$2.6 \times 10^{35} {\rm erg s^{-1}}$, which is roughly 2000 times the
maximum luminosity possible by energy release due to spin down.

Interestingly, \0142 exhibited some bursting activities which were
discovered during the RXTE monitoring of this source. A first burst
was found on April 6, 2006 \citep{Kaspi06_0142atel}. Soon thereafter,
on June 25, four more bursts were detected
\citep{Dib06_0142atel}. During these last events, the source seemed to
be in an active state. Furthermore, \citet{Gavriil07_atel0142}
discovered a very large X-ray burst on February 7, 2007. This burst
was extraordinary as it showed characteristics similar to a SGR giant
flare, only less energetic. It started with a bright and short spike
with a peak flux of $\sim$$2.5 \times 10^{-8}$ \ecs (2--60 keV) and in
the first part of the decaying tail single pulsations of the AXP were
visible. The extended tail lasted up to half an hour. The fluence
($T_{90}$) of this burst was $\sim$$14.5 \times 10^{-8}$ \ecs (2--60
keV), which is 6--88 times higher than the previous five bursts from
\0142. Moreover, the spectrum shows significant spectral features
\citep{Gavriil07_0142bursts}.

Theoretical attempts to find physical explanations for the hard X-ray
emission have started, but so far none of the models can explain all
of the published properties and for some aspects the data cannot
discriminate between the proposed models. More detailed observational
results are required to further constrain the geometries of the
production sites as well as allowed production processes. In the
discussion (see, Sect.~\ref{sec:disc}) we summarise the different
theoretical approaches, the promising progress, but also their
shortcomings.

In this work detailed hard X-ray characteristics are presented using
all available \igr\, data for \0142 to date.  These include total
spectra (pulsed + DC emission) both time-averaged (whole mission) as
well as in 5 shorter time stretches to look for possible long-term
time variability. We also analysed two Target of Opportunity observations
performed shortly after the above mentioned burst activities to look for
variability of any kind. Furthermore, the time-averaged
total-pulsed spectrum as well as time-averaged
phase-resolved pulsed spectra are presented.

In order to study possible relations between the hard ($>$10 keV) and
soft ($<$10 keV) X-ray bands we have (re)analysed archival \xmm,
\pca\, and \asca\, data applying parameters and selections consistent
with the \igr\, analyses.

\section{Observations and analysis}
\label{sec:obs}

\subsection{INTEGRAL}
\label{sec:igr}

The hard X-ray/soft gamma-ray (3 keV -- 8 MeV) mission {\em
  INTErnational Gamma-Ray Astrophysics Laboratory} INTEGRAL
\citep{Winkler03_igr}, has been operational since October 2002.  For
this work both main instruments IBIS and SPI are used.  IBIS, {\em
  Imager on Board the INTEGRAL Satellite} \citep{Ubertini03_ibis}, is
a coded-mask instrument with a low-energy detector called {\em
  INTEGRAL Soft Gamma-Ray Imager} \citep[ISGRI; ][]{Lebrun03_ISGRI}
which is sensitive between $\sim$20~keV and $\sim$300~keV and has a
wide field of view (FOV) of $29^\circ \times 29^\circ$ (full-width
zero response).  The angular resolution is about 12\arcmin.  The {\em
  Spectrometer for INTEGRAL} \citep[SPI; ][]{Vedrenne03_SPI} is also a
coded-mask instrument which is sensitive between $\sim$20~keV and
$\sim$8~MeV. The FOV is $35^\circ \times 35^\circ$ (full-width zero
response). This instrument is optimised for spectroscopy and has a
modest angular resolution of $\sim$2\fdg5. SPI is of particular use in
this work for its better sensitivity to IBIS above $\sim$200 keV.

In its default operation modes, INTEGRAL observes the sky in dither
patterns \citep{Jensen03_igr}. Typical pointings (Science Windows,
ScWs) have integration times of
$\sim$1800--3600~s. Table~\ref{tab:obs} lists the total exposure time
for the selected ScWs for which \0142\, was within an angle of 14\fdg5
from the pointing direction of IBIS, as well as the effective
on-source exposure, which is reduced due to off-axis viewing
angles. Due to the visibility windows of this part of the Galaxy for
INTEGRAL operations, the observations of the prime targets form groups
of consequentive orbital revolutions (Revs).

There are  three dedicated observations on \0142. The
first one was performed mid 2005 and lasted for 1~Ms and two were
shorter Target of Opportunity (ToO) observations which were triggered
by the bursting activities detected with RXTE on June 25, 2006 and
February 7, 2007.

The data were screened for Solar flares and erratic count-rates due to
the passages through the Earth's radiation belts. After screening 1671
ScWs, resulting in 4.7 Ms total exposure time, were available for the
analyses.

\begin{table*}[!t]
\centering
\renewcommand{\tabcolsep}{1.7mm}
\caption[]{Phase-coherent ephemerides derived from RXTE-PCA
monitoring data, valid for the analysed INTEGRAL observations.  }

\begin{tabular}{l l l l l l l r@{.}l l}

\vspace{-3mm}\\

\hline
\hline
\vspace{-3mm}\\
AXP &
Start &
End &
INTEGRAL &
$t_{\rm Epoch}$ &
$\nu$ &
$\dot{\nu}$ &
 \multicolumn{2}{l}{$\ddot{\nu}$} &
$\Phi_0$ \\

 & [MJD] & [MJD] & range (Revs) & [MJD, TDB] &[Hz] & 
 $\times 10^{-13}\,\rm{[Hz\,s^{-1}]}$&  
 \multicolumn{2}{l}{$\times 10^{-22}\,\rm{[Hz\,s^{-2}]}$}& \\
\hline
\vspace{-3mm}\\
\0142 & 52549 & 53169 & 001--203 & 52726 & 0.1150945693297
 & $-$0.267038 &  0&220 & 0.9442\\
\0142 & 53251 & 53619 & 230--353 & 53420 & 0.1150929854501
 & $-$0.263899 &  0&307 & 0.8906\\
\0142 & 53562 & 53745 & 334--396 & 53650 & 0.1150924611500
 & $-$0.267430 & $-$0&319 & 0.2817\\

\hline
\end{tabular}

\label{tab:eph}
\end{table*}

\subsubsection{INTEGRAL spectra from spatial analysis}
\label{sec:igrspec}

For spectral analysis the IBIS-ISGRI data of all 1671 ScWs are reduced
using the Off-line Scientific Analysis (OSA) software package version
5.1 \citep[see][]{Goldwurm03_osa}. For each ScW deconvolved sky images
are created in 20 energy intervals using exponential binning between
20 keV and 300 keV. For every energy band time averaged count rates
are determined by averaging the count rates, weighted by the variances,
from all deconvolved maps at the position of \0142, for each set of
consecutive observations and the total data set. To convert the count
rates to flux values we have used the procedure described by
\citet{Kuiper06_axps}.  The measured AXP count rates are normalized to
the Crab count rates, determined from Crab observations performed
during INTEGRAL Rev.~102. The AXP fluxes in Crab units are converted
into photon fluxes using a curved power-law shape; 

\begin{equation}
F_{\gamma} = 1.5703(14)
\times (E_{\gamma}/0.06335)^{-2.097(2)-0.0082(16)\times {\rm
ln}(E_{\gamma}/0.06335)}
\end{equation}
\citep[Eq.~2 in][]{Kuiper06_axps}, where
$F_{\gamma}$ is expressed in ph/(cm$^2$\,s\,MeV) and $E_{\gamma}$ in
MeV. Finally the spectra are imported in XSPEC version 12.3
\citep{Arnaud96_xspec}, which is used for the spectral fitting. All
errors quoted in this paper are 1$\sigma$ errors, unless stated
otherwise.

The spectral analysis of SPI data is based on all (screened) data up
to Rev.~336.  The data were binned in 9 energy bins covering the
20~keV -- 1~MeV energy range. For each of the energy bins, source
fluxes have then been extracted using a maximum likelihood fitting
procedure that considers the instrumental response of the telescope
\citep{Knoedlseder04_spi}. The measured flux is then normalized to the
Crab spectrum that has been obtained by fitting the SPI data of
revolutions 43--45. Multiplication with the above-mentioned curved
power-law Crab model results then into photon fluxes.

\subsubsection{INTEGRAL timing analysis}
\label{sec:igrtiming}

For the INTEGRAL timing analysis, we have followed the procedure
described by \citet{Kuiper06_axps}. We only considered ISGRI data for
this purpose. Data up to Rev.~396 are used (see Table~\ref{tab:obs}).
Events are selected from non-noisy detector pixels that were/could be
illuminated by the source through an open mask element with a greater
illumination factor than 25\%. Instrumental, on-board processing and
ground-station time delays are corrected for \citep{Walter03_timing}.
The event arrival times at the space craft are converted to arrival
times at the Solar-system barycenter using the JPL DE200 Solar-system
ephemeris and the source position measured with {\it Chandra}
\citep{Juett02_0142}. The barycentered arrival times are finally
folded on an appropriate phase-connected ephemeris (see Table
\ref{tab:eph}) created using contemporaneous RXTE data (see
Sect.~\ref{sec:rxte}).  The (TDB) time to pulse phase conversion
taking into account consistent phase alignment for each ephemeris is
provided by the following formula:
\begin{equation}
\Phi(t) = \nu \cdot (t-t_{\rm Epoch}) + \frac{1}{2}\dot{\nu}\cdot
(t-t_{\rm Epoch})^2 + \frac{1}{6}\ddot{\nu}\cdot (t-t_{\rm Epoch})^3 -
\Phi_0.
\end{equation}
The detection significances of the derived pulse profiles are estimated 
applying the Z$^2_{{\rm n}}$ test \citep{Buccheri83_zn2}.

In order to derive the spectrum of the pulsed emission, first the
number of excess counts in the obtained pulse profiles above flat
background/DC levels have to be determined. For this purpose, the pulse
profiles for the selected energy bands are fitted with truncated
Fourier series using two harmonics. The minimum of this fit is defined
as the DC level. The excess counts are counted above this
level and converted into flux units applying the same procedure for
the Crab pulsar and normalising to the known Crab-pulsar spectrum
\citep[Eq.~3 in][]{Kuiper06_axps};
\begin{equation}
F_{\gamma} = 0.4693(21)
\times (E_{\gamma}/0.04844)^{-1.955(7)-0.0710(78)\times {\rm
ln}(E_{\gamma}/0.04844)}.
\end{equation}
Phase-resolved spectra are similarly created using only excess counts
within selected phase intervals (see Sect.~\ref{sec:phaseres}).


\subsection{RXTE}
\label{sec:rxte}

The {\em Rossi X-ray Timing Explorer} (RXTE) has been operational
since 1996. For this work we use data from the Proportional Counter
Arrays (PCA) onboard RXTE \citep{Jahoda96_pca}, which is a non-imaging
instrument sensitive in the 2--60 keV energy range. The FOV is
approximately 1$^{\circ}$ (FWHM).

Regular monitoring observations of \0142\, with the \pca\,
\citep{Gavriil02_rxtemonitoring} are the basis of the timing results
in this paper. These observations allow us to create accurate
phase-connected timing solutions. We have generated three new
phase-connected ephemerides with consistent pulse alignment to cover
the whole INTEGRAL-time span.  The details of the ephemerides can be
found in Table~\ref{tab:eph}. They are in agreement with the solution
presented by \citet{Dib07_0142rxte}, who also present a phase-coherent
timing solution over the period 1996--2006.

We have used the same \pca\, data as listed in Table~1 of
\citet{Kuiper06_axps} which were taken between March 1996 and
September 2003 before the \igr\, observations, to perform spectral
timing analyses analogous to the procedures outlined by
\citet{Kuiper06_axps}.  We note that in the folding process of PCA
data the application of ephemeris entry one of Table~\ref{tab:eph}
results in pulse profiles which are shifted 0.543 in phase with
respect to the profiles shown in \citet{Kuiper06_axps} for \0142.

Irrespective of the energy band, the pulse-profiles can be described
sufficiently accurate by truncated Fourier series with 3 harmonics
above constant background levels. The excess (pulsed) counts above the
background levels are converted into flux units using properly PCU
exposure weighted energy response matrices \citep[see e.g. Sect.~3.2
  of][]{Kuiper06_axps}.

We have extracted a time-averaged total-pulsed spectrum and
phase-resolved spectra adopting the same phase intervals as selected
for \igr\, (see Sect.\ref{sec:phaseres}). To derive unabsorbed (soft)
X-ray spectra a Galactic Hydrogen absorption column of $N_{\rm H} =
0.57 \times 10^{22}$ cm$^{-2}$ has been used (as fixed value) as
determined from spectral analyses of \xmm\, observations (see
Sect.~\ref{sec:xmmtot}).


\subsection{XMM-Newton}
\label{sec:xmm}

\begin{table}[!htb]
\centering
\renewcommand{\tabcolsep}{1.7mm}
\caption[]{XMM-Newton observations of \0142. The observations are
  designated A, B and C and referenced accordingly in the
  text. Further, the observation dates, exposures and Good Time
  Interval (GTI) exposures for MOS and PN are given.}

\begin{tabular}{llllll}

\vspace{-3mm}\\

\hline
\hline
\vspace{-3mm}\\
$\#$ & Obs ID & Date & Exp. & GTI-MOS & GTI-PN\\
     &        &      & (ks) & (ks)    & (ks)\\

\hline
\vspace{-3mm}\\
A & 0112781101 & 2003-01-24 &  6.4 &  5.4 &  3.7\\
B & 0206670101 & 2004-03-01 & 46.5 & 36.4 & 27.3\\
C & 0206670201 & 2004-07-25 & 23.9 & 23.0 & 20.6\\

\hline
\end{tabular}

\label{tab:xmm}
\end{table}
\begin{figure*}
\centering
\begin{minipage}[c]{0.8\textwidth}
\psfig{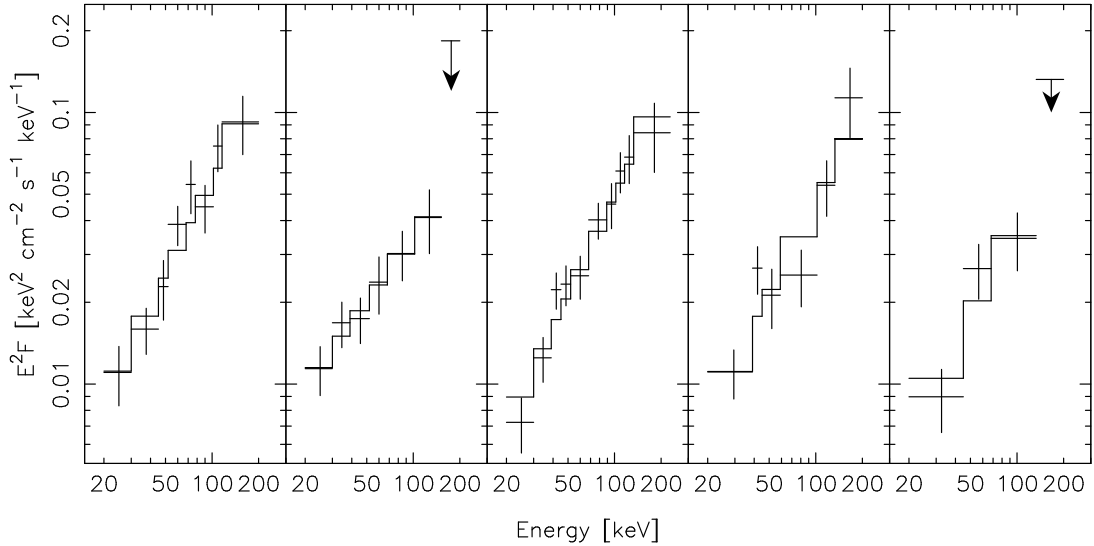}
\end{minipage}
\caption{Total-flux spectra of 4U0142+61 taken during
  \igr\ revolutions (from left to right); 142--189, 202--269,
  331--336, 384--396 and 452--468. Data points are binned to $4\sigma$
  if the single bins are of less significance. The upper limits
  represent 2$\sigma$ confidence levels.
\label{fig:5spec}}
\end{figure*}

\xmm\, \citep{Jansen01_xmm} has been operational since 2000. Onboard
are three CCD cameras for X-ray imaging, namely two EPIC (European
Photon Imaging Camera) MOS cameras \citep{Turner01_mos} and one
EPIC-PN camera \citep{Strueder01_pn}. All cameras have a FOV of
$\sim$30\arcmin\, and are sensitive in the energy range $\sim$0.3--12
keV.

We have (re)analysed the four publicly available data sets acquired by
\xmm\, between 2002 and 2004. The data set taken on February 13, 2002
(Obs Id 0112780301) did not contain usable data for this work. The
summary of the remaining 3 data sets can be found in
Table~\ref{tab:xmm}. We designate these observations A, B and C,
respectively. In this work we present only the results using the
PN. We have also analysed the data from both MOS instruments and used
this for consistency checks.  The PN was in small-window mode during
observation A and in timing mode during observations B and C. In
parallel studies \citet{Rea07_0142xmm} and \citet{Gonzalez07_0142}
also analysed these \xmm\ observations. \citet{Gonzalez07_0142} also
included private \xmm\ data taken in 2006 and 2007 together with {\em
  Chandra} and {\em Swift} data to study long-term variations below
10~keV.


The three data sets in Table~\ref{tab:xmm} have been analysed using
SAS v. 7.0 and the latest calibration files that were available
(Jan. 2007). 
The data have been checked for solar (soft proton) flares by creating
a light curve for each observation for events with energies larger
than 10~keV. From these light curves the count-rate distribution has
been determined and subsequently fitted with a Gaussian. Good Time
Intervals (GTIs) have been created by allowing only time stamps from
time intervals for which the count rate was lower than the fitted mean
count rate plus three times the width of the distribution
(3$\sigma$). For observations A, B and C these cutoff values were
0.082, 0.243 and 0.358 counts per second. PN single and double events,
i.e. patterns less than and equal four, and an energy range of 0.3 to
12.0~keV have been selected. The arrival times of the selected events
are barycentered adopting the most accurate celestial position of
\0142 \citep{Juett02_0142}.

There was no need to correct for pile up in observation A as the count
rate from \0142 was sufficiently low for the small-window mode. A
circular extraction region centered at \0142\, with a radius of
40\arcsec\ was used.  For the background extraction two circular
source-free regions centered at different locations in the
small-window were chosen with radii of 35\arcsec\ and 47\farcs5,
respectively.

Observations B and C were taken in timing mode. For these observations
we chose extraction regions with a total width of 27 pixels centered
on the source, which corresponds to 110\farcs7 close to the 95\%
encircled energy radius of the source.  While the sensitive area (ARF)
generator corrects for the remaining $\sim$5\% of missed energy
outside the source extraction region, it does not compensate for the
subtraction of any source contribution in the chosen background
extraction strip.  We took a strip of 11 pixels wide starting
102\farcs5 from the point-spread function (PSF) core. The source
contribution from this part of the PSF can be neglected. We fitted a
Lorentzian plus constant background to the observed one-dimensional
source profile.  The source contribution in the 11 pixel wide
background strip turned out to be a mere 0.48\% of the normalized PSF,
which means taking into account the source-to-background scaling of
27/11 an ignored correction of about 1.2\%.

The extracted spectra have been binned, oversampling the energy
resolution by a factor of three and then occasionally rebinned to
ensure a minimum of 25 counts per bin.

For the \xmm\ timing analysis, the events selected as described above
are used. These events are folded using the appropriate ephemeris in
Table~\ref{tab:eph}. For the pulsed spectra the pulsed excess counts are
extracted using the same procedure as described in
Sect.~\ref{sec:igrtiming} for the \igr\ pulsed spectrum, fitting the
pulse profiles with truncated Fourier series with the first three
harmonics above a constant background.


\subsection{ASCA}
\label{sec:asca}

ASCA \citep{Tanaka94_asca} was operational for eight years from
1993. Onboard ASCA was the {\em Gas Imaging Spectrometer}
\citep[GIS,][]{Ohashi96_gis, Makishima96_gis}. The GIS was sensitive
between $\sim$0.7--12~keV and had a FOV of $\sim$50\arcmin.

We have revisited the spectral timing analysis by
\citet{Kuiper06_axps} of the \asca\ observation of \0142 taken in
July/August 1999 to make it consistent with this work, i.e., in the
current spectral analysis we have changed and fixed the Galactic
absorption column to the value derived from the \xmm\ data (see
Sect.~\ref{sec:xmmtot}). In the timing analysis the pulse phase has
been redefined to be consistent with that of \igr, RXTE and \xmm,
allowing the same phase selections for the phase-resolved
spectroscopy.


\section{Results}
\label{sec:results}

In this section, we will first present the \igr\ and \xmm\ total
spectra and the search for long-term variability. Next, we will
present pulse profiles from our timing analyses of \igr, \xmm, RXTE
and ASCA. Finally, for the same instruments the total pulsed and
phase-resolved spectra are derived consistently, revealing that
genuinely different components contribute to the total pulsed
emission.


\subsection{Total spectrum and long-term variability}
\label{sec:totspec}

Previous INTEGRAL studies by
\citet{denHartog06_casa,denHartog07_london} and \citet{Kuiper06_axps}
show that hard X-rays from \0142 have been detected at different
epochs at similar flux levels, suggesting a stable and persistent hard
X-ray spectrum. Firstly, we will verify the stable nature of the
high-energy emission. Then, we will present the time-averaged total
spectrum using contemporaneous IBIS-ISGRI and SPI data.  Finally, the
total spectra of \xmm\, are derived and an example is shown together
with the \igr\, time-averaged total spectrum, rendering a broad-band
X-ray view of the total spectrum.

\subsubsection{INTEGRAL ISGRI persistent spectra and long-term variability}
\label{sec:igrvar}

\begin{figure}
\psfig{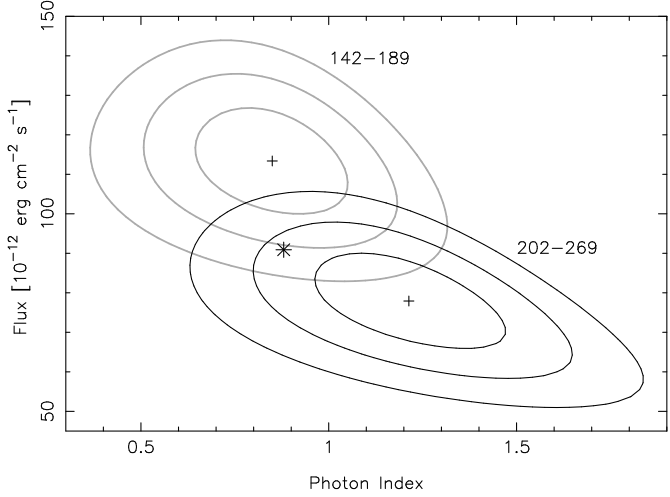}
\caption{Error contours ($1\sigma, 2\sigma$ and $3\sigma$) of
  observations 142--189 (in grey) and 202--269 (in black) with respect
  to the time-averaged measurement (star).
\label{fig:contours}}
\end{figure}

The ISGRI data were grouped in five sets as indicated in
Table~\ref{tab:obs} and analysed as described in
Sect.~\ref{sec:igrspec}.  The statistical quality is not equal for
every spectrum, but \0142\, is detected significantly in each group
and spectra can be extracted from each data set. The resulting five
total spectra are displayed in Fig.~\ref{fig:5spec}.

To fit the spectra we use single power-law models. As in previous work
they describe the spectral shapes at a satisfactory level. The fit
results for these spectra are listed in the upper part of
Table~\ref{tab:fits}. The photon indices $\Gamma$ range from $0.79 \pm
0.10$ to $1.21 \pm 0.16$ and the fluxes in the 20--150 keV energy band
($F_{20-150}$) range from $7.8 \times 10^{-11}$ to $11.3 \times
10^{-11}$ \ecs.

The errors in the indices $\Gamma$ and fluxes $F_{20-150}$ are
  not independent. In order to  check for possible long-term time
variability we therefore compared the error contours of the
different fits with respect to the error contours of the fit to the
time-averaged total spectrum (see below in
Sect.~\ref{sec:igrtot}). The spectra extracted for Revs. 142--189 and
202--269  are the most deviating with respect to the time-averaged
  spectrum. In Fig.~\ref{fig:contours} the error contours for these
spectra are shown together with the best-fit values for the
time-averaged total spectrum (Revs. 142--468).  The two spectra
deviate from the average around the $\sim$2$\sigma$ level.

The standard deviations ($s = \sqrt{ 1/(n-1) \sum_i (x_i - \bar{x})^2
}$) for the power-law indices and the 20--150 keV fluxes relative to
the weighted means are 0.16 and $1.5 \times 10^{-11}$ \ecs,
respectively. Therefore, both the power-law shape and the 20--150 keV
flux of \0142 are stable within 17\% (1$\sigma$).

\begin{table}[t]
\centering
\renewcommand{\tabcolsep}{1.7mm}
\caption[]{Summary of the  single power-law fits for the
  INTEGRAL-ISGRI total-spectra of \0142 at different epochs for the
  observations summarized in Table~\ref{tab:obs}. For each epoch the
  photon index, 20--150 keV flux and \chir\, are given. For the
    the spectrum taken in Rev. 142--468 also the 20--229~keV flux is
    given.}

\begin{tabular}{l r@{ $\pm$ }l r@{ $\pm$ }l l }

\vspace{-3mm}\\

\hline
\hline
\vspace{-3mm}\\
Rev.&
\multicolumn{2}{c}{$\Gamma$}& 
\multicolumn{2}{c}{$F_{20-150} \times 10^{-11}$} & 
\chir\ (dof)  \\

 &  \multicolumn{2}{c}{}  & \multicolumn{2}{c}{[\ecs]}\\
\hline
\vspace{-3mm}\\
142--189 & 0.85&0.14 & 11.34&0.78 & 0.69 (15) \\
202--269 & 1.21&0.16 &  7.80&0.83 & 0.66 (13) \\
331--336 & 0.79&0.10 &  9.80&0.57 & 1.12 (16) \\
384--396 & 0.89&0.17 &  9.46&0.91 & 0.91 (15) \\
452--468 & 0.96&0.25 &  7.8 &1.2  & 0.82 (12) \\
\hline
\vspace{-3mm}\\
454 ToO  & 1.06&0.30 &  7.8&1.4   & 0.98 (12) \\
528 ToO  & 0.88&0.34 &  9.3&1.8   & 1.63 (11) \\
\hline
\vspace{-3mm}\\
142--468 & 0.93&0.06 &  9.09&0.35$^\dagger$ & 0.94 (16)\\
 \multicolumn{6}{l}{$^\dagger F_{20-229} = (15.01 \pm 0.82) \times 10^{-11}$ \ecs}\\
\hline
\end{tabular}

\label{tab:fits}
\end{table}

\subsubsection{INTEGRAL ISGRI \& SPI total spectra}
\label{sec:igrtot}

\begin{figure}
\psfig{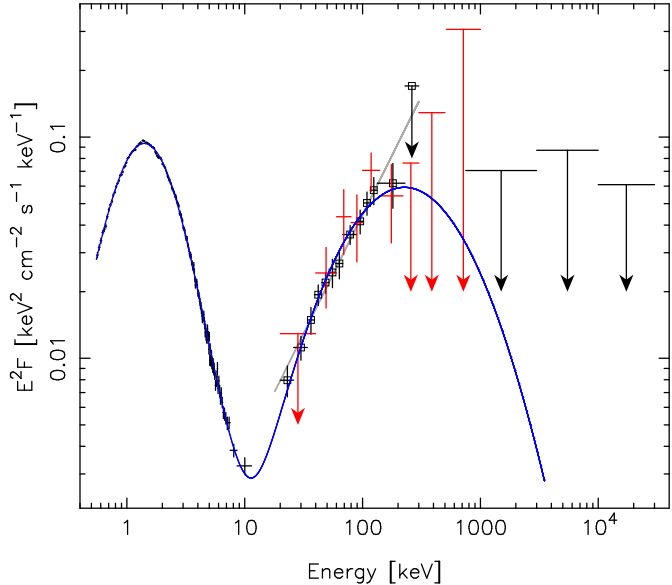}
\caption{Unabsorbed total spectra of 4U~0142+61 as measured with
  different instruments; XMM-Newton (Obs B; 0.55--11.5 keV) in black;
  INTEGRAL-ISGRI (Revs. 142--468; 20--300 keV) in black with open
  square symbols; INTEGRAL SPI (20--1000 keV) in red; and CGRO COMPTEL
  (0.75--30 MeV) limits in black. Shown are also the best single
  power-law model fit for the ISGRI data-points (in grey) and the
  `three logparabola' fit for the whole band (in blue). See
  Sect.~\ref{sec:totspec} for details.
\label{fig:tothigh}}
\end{figure}

Using all available (Revs.~142--468, see Table \ref{tab:obs}) 2.37~Ms
net on-source exposure the time-averaged total spectrum
(Fig.~\ref{fig:tothigh}) derived from ISGRI data can be fitted
satisfactorily with a single power-law model. The best fit parameters
give a photon index $\Gamma = 0.93 \pm 0.06$ and a 20--150 keV flux of
$(9.09 \pm 0.35) \times 10^{-11}$ \ecs.  \0142\, has been detected up
to 229 keV with a 4.3$\sigma$ significance in the 152--229 keV
band.

Even though the upper limits at MeV energies obtained with the Compton
telescope COMPTEL onboard CGRO \citep[][ also shown in
  Fig.~\ref{fig:tothigh}]{denHartog06_casa} strongly suggest that the
spectrum has to break at a few hundred keV, the ISGRI spectrum shows
no indication for a spectral break. The quality of the fit (\chir\, =
0.94; dof = 16) assuming a single power-law model over the 20-300 keV
range is good and the use of a more complex model with additional
parameters is statistically not justified.

SPI is more sensitive than ISGRI at energies higher than $\sim$200
keV. Therefore, we have also analysed all available SPI data with
\0142 in the FOV. We have extracted the SPI spectrum and detected
\0142\, significantly above 40 keV up to 140 keV. The maximum
detection significance is 5$\sigma$ between 80 keV and 140 keV. In the
next energy bin (140--220 keV) a marginal detection of 2.6$\sigma$ is
achieved. The six flux values between 20 and 220 keV can be fitted
(\chir\, = 1.67; dof = 4) with a power-law model with photon index
$\Gamma = 0.86 \pm 0.20$ and a 20--150 keV flux of $(8.8 \pm 1.2)
\times 10^{-11}$ \ecs, which is consistent with the ISGRI total
spectrum (see Fig.~\ref{fig:tothigh}).

Fitting the SPI spectrum with a single power law using all 9
spectral bins between 20 and 1000 keV we find a poor and
unsatisfactory fit with a \chir\ = 2.57 for 7 dof ($\chi^2 = 18$). 
Therefore, we followed \citet{Kuiper06_axps}, by using a logparabolic
function;
\begin{equation}
\label{eq:lp}
F = F_{0} \times \left(\frac{E}{E_{0}}\right)^{-\alpha -\beta
  \cdot{\rm ln}\left(\frac{E}{E_0}\right)}
\end{equation}
with $E_0$ (in units keV) the pivot energy introduced to minimize
correlations between the parameters, and $F_0$ the flux (in units
\pcsk) at $E_0$. Note that this function is a simple power law if the
curvature parameter $\beta$ is equal to zero.

The fit of the SPI spectrum improves with $\Delta \chi^2_1 = 15.77$,
which corresponds to a $\sim 4\sigma$ improvement for one additional
fit parameter.

To exploit all available spectral information we fitted the ISGRI and
SPI spectra simultaneously. Starting from a single power-law model we
obtained a fit with a \chir\ = 1.58 for 27 dof ($\chi^2 = 42.8$)
which can be improved significantly. The model described by this
power-law is too high in the energy range with the SPI upper limits
i.e. above 220 keV. Using again a logparabolic function an optimum fit
with a $\chi^2$ of $20.1$ is achieved. A $\Delta \chi^2_1$ of 22.7 for
one additional parameter translates in a 4.8$\sigma$ fit
improvement. This is the {\em first} clear detection of the spectral
break in the total spectrum of \0142\ above 20~keV using
contemporaneous high-energy data.  The best fit parameters are $\alpha
= 1.26 \pm 0.09$, $\beta = 0.41 \pm 0.09$ and $F_0 = (5.1 \pm 0.3)
\times 10^{-6}$ \pcsk\ at $E_0$ = 91.527 keV.  The peak energy (for a
$\nu F_\nu$ representation, eq. to $E^2F$) corresponding to these
parameters is $E_{\rm peak} = 228^{+106}_{-53}$ keV, while the 20--150
keV flux amounts: $(8.97 \pm 0.86) \times 10^{-11}$ \ecs\ (fitting
three free parameters).

Including the non-contemporaneous COMPTEL (0.75--30 MeV) flux
measurements (upper limits in Fig.~\ref{fig:tothigh}) in a similar fit
yielded best fit parameters with somewhat smaller uncertainties.  We
found as best fit parameters: $\alpha = 1.484 \pm 0.057$, $\beta =
0.351 \pm 0.044$ and $F_0 = (3.01 \pm 0.16) \times 10^{-6}$ \pcsk\ at
$E_0$ = 133.718 keV.  The peak energy ($E^2F$) corresponding to these
parameters is $E_{\rm peak} = 279^{+65}_{-41}$ keV.  We note that this
result is based on an assumed ``logparabolic" spectral shape.  We
adopted this spectral form in order to comply with the SPI and COMPTEL
limits. However, we do not claim that we have measured the spectral
shape above $\sim$200~keV in detail. In particular, the extension of
the spectrum towards MeV energies is uncertain. Significant
measurements in the energy range between 200~keV and 1~MeV are
required to unravel the true spectral shape in the vicinity of the
break energy.

\subsubsection{\igr\ post-burst spectra}
\label{sec:igrtoo}

Of the five AXPs for which bursts have been detected, \0142 is so far
the only one for which a hard X-ray spectrum above $\sim$10~keV
has been observed. After two remarkably different bursting events 
detected with \pca\ we requested two \igr\ ToO observations in order to
study possible correlations between the soft- and hard X-ray emission.
For both occasions $\sim$200~ks ToO time was granted. The source was 
observed in Revs. 454 and 528 (see Table~\ref{tab:obs}).

We have detected \0142\, with detection significances in the 20-150
keV energy band of 5.4$\sigma$ and 5.9$\sigma$ for Rev-454 and
Rev-528, respectively.  Both spectra have been fitted with a power-law
model which described the spectra well. The fit parameters for the
first ToO are $\Gamma = 1.06 \pm 0.30$ and $F_{20-150} = (7.8 \pm 1.4)
\times 10^{-11}$ \ecs and for the 2nd ToO the fit parameters are
$\Gamma = 0.88 \pm 0.34$ and $F_{20-150} = (9.3 \pm 1.8) \times
10^{-11}$ \ecs.  The values are fully consistent with those for the
time-averaged spectrum\footnote{Note that the 1st ToO is included in
  the time-averaged total spectrum, but we have tested the ToO with
  the time-averaged total spectrum of the first four data sets from
  which it was not statistically different.}  (see also
Table~\ref{tab:fits}).

\citet{Gonzalez07_0142} reported a flux increase of (15$\pm$3)\% in
the 2--10~keV band observed with \xmm\ coinciding with the burst
activity in 2006--2007. It should be noted, however, that in order to
measure with \igr\ a significant ($>$3$\sigma$) change in the hard
X-ray spectrum within a 200~ks observation, the flux level should have
changed by at least $\sim$60\%.  Correlated changes in  the hard and
soft X-ray fluxes of the scale reported by
\citet{Gonzalez07_0142} could therefore not be measured.

\subsubsection{\xmm\ EPIC-PN total emission spectra}
\label{sec:xmmtot}
\begin{table*}
\centering
\renewcommand{\tabcolsep}{1.7mm}
\caption[]{XMM-Newton EPIC-PN spectral fits of the total-emission
  spectrum of \0142. Three logparabolae (Eq.~\ref{eq:lp}) were used
  for an empirical fit including three fixed parameters from the
  ISGRI/SPI 20--1000 keV spectral fit. For each observation (A, B and
  C, see Table~\ref{tab:xmm}) six optimum parameter values are shown,
  three for each logparabolic model (indicated with a subscript
  number), for the best fit (0.55--10 keV).  The last row lists the
  unabsorbed 2--10~keV energy flux.}

\begin{tabular}{lccc}
\vspace{-3mm}\\
\hline
\hline
\vspace{-3mm}\\
  & A & B & C \\
\hline
\vspace{-3.0mm}\\
$F_{0,1}$          & ($6.85   \pm  0.05)  \times  10^{-2}$
               & ($5.50   \pm  0.02)  \times  10^{-2}$
               & ($4.57   \pm  0.02)  \times  10^{-2}$\\
$\alpha_1$     & $0.933  \pm  0.012$
               & $0.673  \pm  0.006$
               & $0.528  \pm  0.008$\\
$\beta_1$      & $1.534  \pm  0.013$
               & $1.899  \pm  0.007$ 
               & $2.089  \pm  0.010$\\
\vspace{-2.0mm}\\
$F_{0,2}$          & ($1.04  \pm  0.04)  \times  10^{-2}$
               & ($2.33  \pm  0.02)  \times  10^{-2}$
               & ($3.20  \pm  0.02)  \times  10^{-2}$\\
$\alpha_2$     & $2.095 \pm 0.041$
               & $1.671 \pm 0.009$ 
               & $1.440 \pm 0.008$\\
$\beta_2$      & $0.212 \pm 0.025$
               & $0.622 \pm 0.007$ 
               & $0.808 \pm 0.006$\\
\vspace{-2.0mm}\\
$F_{2-10\,\rm{keV}}$ (\ecs) & $(6.68 \pm 0.40) \times 10^{-11}$
                           & $(6.27 \pm 0.14) \times 10^{-11}$ 
                           & $(6.49 \pm 0.15) \times 10^{-11}$ \\
\vspace{-3.0mm}\\ 

\hline
\vspace{+0.0mm}\\ 
\multicolumn{4}{l}{a) The photon fluxes $F_{0,1}$ and
  $F_{0,2}$ are expressed in \pcsk at $E_0$ = 1 keV.}\\
\multicolumn{4}{l}{b) Error estimates ($1\sigma$) in the fit
  parameters are obtained by fixing all parameters except the
  parameter}\\
\multicolumn{4}{l}{\ \ \ \, of interest at their optimum values and
  subsequently determining, where the parameter of interest}\\
\multicolumn{4}{l}{\ \ \ \, reaches a $\Delta\chi^2_{6-1}$ step of
  7.038 from the global minimum.  This increment corresponds to a
  $1\sigma$ step}\\
\multicolumn{4}{l}{\ \ \ \, for 6$-$1 degrees of freedom.}\\
\end{tabular}

\label{tab:xmmfits}
\end{table*}

We derived the \xmm\ EPIC-PN total (= pulsed plus DC) spectra as
described in Sect.~\ref{sec:xmm} (see e.g. Fig~\ref{fig:tothigh} for
an example).  We first investigated the absorption column ($N_{\rm
  H}$) using the XMM observations with the best statistics (B and C)
and concentrated ourselves on energies below $\sim$5~keV i.e. the soft
part of the X-ray spectrum. 

We used logparabolic functions in order for the spectra to bend
downwards to the optical regime, as required by the  broad-band
spectrum \citep[see e.g. the spectrum from the first  broad-band
  campaign by][]{denHartog07_london}.  If a traditional model, a
combination of a black body and a power law, was chosen, the soft
power-law component would keep increasing towards lower energies,
which can only be compensated by increasing the absorption column.
The latter would be forced to unrealistically high values \citep[see
  e.g.][ who derived $N_{\rm H} = (1.00 \pm 0.01) \times 10^{22}$
  cm$^{-2}$]{Rea07_0142xmm}. Fixing the $N_{\rm H}$ to the value
published by \citet[][ $N_{\rm H} = (0.64 \pm 0.07) \times 10^{22}$
  cm$^{-2}$]{Durant06_extinction} from an accurate analysis of data of
the Reflection Grating Spectrometer aboard \xmm\ yielded, however,
also poor fits at low energies between $\sim$0.6~keV and $\sim$1~keV
in both observations B and C. Therefore, we treated $N_{\rm H}$ as a
free parameter. For both observations we find an absorption column of
$(0.57 \pm 0.02) \times 10^{22}$~cm$^{-2}$, which is somewhat lower
than, but consistent with the value obtained by
\citet{Durant06_extinction}.  Therefore, assuming that the origin of
the absorption is Galactic and thus constant, we fix $N_{\rm H}$ to
$0.57 \times 10^{22}$ cm$^{-2}$ for the rest of the \xmm\ analyses.
We note that \citet{Guver08_0142} find similar values for $N_{\rm
    H}$ by applying a surface thermal-emission and magnetosphere
  scattering model to the same \xmm\ data in a parallel analysis.

Next, we consider the total \xmm\ spectral energy window up to
$\sim$12~keV.  An additional model component is required to describe
at least a part ($<20\%$) of the harder ($\sim$5--12~keV) X-ray
emission from both observations B and C. For this component a second
logparabolic function is introduced, instead of the traditionally
introduced power-law component. Furthermore, a third logparabolic
component with {\em fixed} model parameters prescribed by the
ISGRI/SPI joint fit above 20 keV (see Sect.~\ref{sec:igrtot})
contributes in this energy band and is included in the total fit.
Without the need to introduce an \xmm-ISGRI inter-calibration factor,
a good fit was obtained to the high-energy parts of the \xmm\ spectra.
In Table~\ref{tab:xmmfits} the fit results are presented for the three
\xmm\ observations.

However, for both long observations (obs. B and C) statistically
unacceptable fits are obtained employing this empirical approach.
Namely, significant residuals are apparent below 2~keV, because the
logparabolae cannot follow the spectral shape exactly at these low
energies. The small deviations from the optimum fit (hardly visible in
Fig.~\ref{fig:tothigh}) are of the order of a few percent for energies
below 2~keV.  We use the spectral fits, however, primarily for the
conversion of absorbed fluxes into unabsorbed fluxes (data points
shown in Fig.~\ref{fig:tothigh}) by calculating for each energy bin
the fraction of energy which is absorbed.  In Sect.~\ref{sec:phaseres}
we use these high-statistics unabsorbed total flux measurements in
even broader bins for deriving the pulsed fraction as a function of
energy, and the uncertainty in that analysis is dictated by the errors
in the fluxes from the phase resolved analysis. Therefore, the
empirical two plus one (fixed) logparabolae fits assuming a constant
absorption column of $0.57 \times 10^{22}$~cm$^{-2}$ are sufficiently
accurate for our purposes.

Note from Table~\ref{tab:xmmfits} that the total unabsorbed 2--10~keV
energy flux is consistent with being constant during the time span of
the three \xmm\ observations. This is in agreement with
\citet{Rea07_0142xmm} and \citet{Gonzalez07_0142}, however,
\citet{Gonzalez07_0142} report a flux increase of (15$\pm$3)\%
(2--10~keV) between the epochs of our \xmm\ observations A, B and C
and the epochs after the bursting activities started in 2006.  The
flux levels around 10~keV of the three \xmm\ spectra are rather
similar.  Fig.~\ref{fig:tothigh}, showing only the \xmm\ obs.~B
spectrum, is therefore a good representation of the total high-energy
($>$0.5~keV) spectrum, particularly showing the abrupt transition from
soft to hard X-rays.


\subsection{Timing analysis: pulse profiles}
\label{sec:timingpp}
\begin{figure}
\psfig{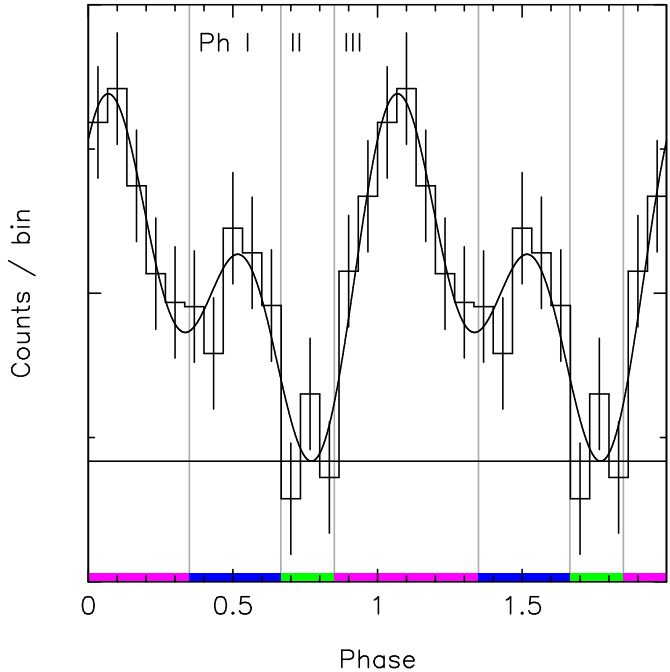}
\caption{20--160~keV ISGRI pulse profile of \0142. The significance is
  $6.5\sigma$ using a ${\rm Z}^2_2$ test (fit shown as solid
  curve). The fitted DC level is indicated with a horizontal line.
  Also indicated are the phase intervals for Ph\,I, II and III (see
  Table~\ref{tab:ph}) with vertical grey lines and with colours.
\label{fig:igrpptot}}
\end{figure}

In this section we first present the time-averaged INTEGRAL IBIS-ISGRI
pulse profiles of \0142. Then, we will examine the \xmm\ EPIC-PN and
\asca\, pulse profiles in single observations and test these for
variability with time and energy, respectively. Finally, we compare
the time-averaged ISGRI pulse profiles with the time-averaged \xmm\,
pulse profiles.

\subsubsection{INTEGRAL ISGRI pulse profiles}
\label{sec:igrpp}

Using the procedure described in Sect.~\ref{sec:igrtiming} it is
possible to detect significant pulsed emission from \0142 in the ISGRI
data. For the energy range 20--160~keV a maximum significance is
reached of $6.5\sigma$ using a ${\rm Z}^2_2$ test. The 20--160~keV
pulse profile, shown in Fig.~\ref{fig:igrpptot}, is broad and appears
to be double peaked with a DC level in a narrow phase interval of
width $\sim$0.2.

Pulse profiles for the differential energy intervals 20--50~keV and
50--160~keV are shown in Fig.~\ref{fig:allpp}\,E,\,F. The significances of
the ISGRI pulse profiles are 4.4$\sigma$ and 4.7$\sigma$,
respectively, and are both more significant than the RXTE-HEXTE
profiles for similar energy ranges (3.4$\sigma$ and 2.0$\sigma$,
respectively) shown in \citet{Kuiper06_axps}. The profile for the
20--50~keV band (Fig.~\ref{fig:allpp}\,E) shows the two peaks separated
$\sim$0.5 in phase.  The 50--160~keV profile (Fig.~\ref{fig:allpp}\,F),
however, shows only a shoulder to the first pulse at the location of
the second pulse (near phase $\sim$0.5) in the 20--50~keV profile.
The minimum (DC level) is in both cases located in the same narrow
phase interval.

In order to quantify a possible pulse-morphology change as a function
of energy, we performed a statistical method based on a combination of
a Pearson $\chi^2$ \citep[see e.g. Sect.~11.2 of][]{Eadie71} and a
Run test \citep[see e.g. Sect.~11.3.1 of][]{Eadie71}, comparing the
shapes of the (independent) 20--50~keV and 50--160~keV pulse profiles.
We did not apply the Kolmogorov-Smirnov test, because these background
dominated (binned) ISGRI pulse phase distributions have different
backgrounds and DC levels. In our approach we first obtained an
appropriate model function describing the shape of the 20--50~keV
pulse profile by fitting a truncated Fourier series using three
harmonics.  Next, we fitted the 50--160~keV distribution (in 20 bins)
in terms of a constant and a function with free scale describing the
shape of the 20--50~keV pulse profile (dof = 18).  We investigated not
only the absolute values of the deviation of the data from the optimum
fit, but also the {\em signs} of the deviations by combining the
probabilities from the independent Pearson $\chi^2$ test
($P_{\chi^2}$) and Run test ($P_{run}$) in an overall joint
probability, $P = P_{\chi^2} \cdot P_{run} (1-\ln{P_{\chi^2}} \cdot
P_{run})$ \citep[see e.g. Sect. 11.6 of][]{Eadie71}.  In our case the
joint probability is 0.104, thus we cannot claim a significant
difference in shape between the 20--50~keV and 50--160~keV ISGRI pulse
profiles.

\subsubsection{\xmm\ and ASCA pulse profiles: time variability}
\label{sec:softpp}

The \asca\ and \pca\ pulse profiles published by \citet{Kuiper06_axps}
suggested that the pulse profiles below $\sim$2--3 keV are
significantly different. However, the \pca\ pulse profiles were
averages over years and the \pca\ is not sensitive below 2~keV,
contrary to \asca\,. \citet{Dib07_0142rxte} used the 10 years of RXTE
monitoring above 2~keV to show long-term variability in the
pulse-profile shape. However, the reported effect is small; mainly the
dip of emission between the two peaks got shallower between 2002 and
2006. Therefore, we used the \asca\ and \xmm\ observations to create
pulse profiles in the same differential energy bands to test for time
variability. The differences in the energy resolutions and responses
between the two instruments has negligible impact on the comparison of
the profile shapes in the selected broad energy intervals.

\begin{figure}
\psfig{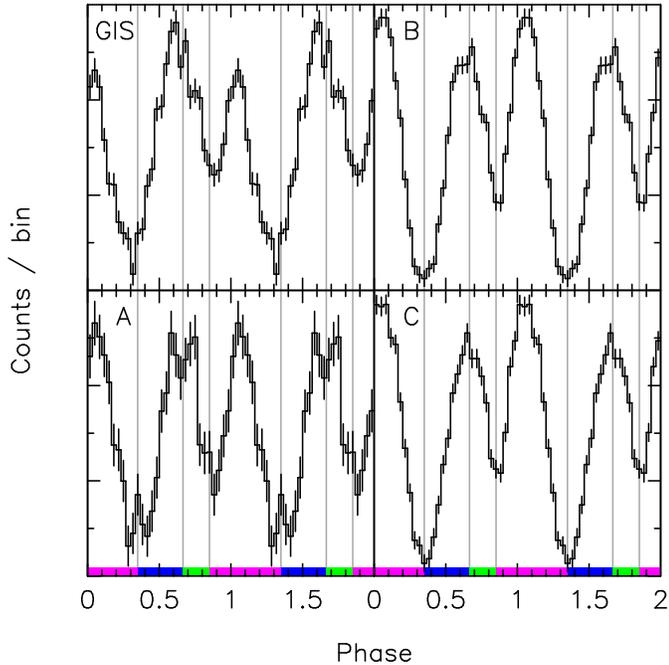}
\caption{\0142\ pulse profiles for 0.8--2.0 keV of \asca\ and the
  three \xmm\ observations (A, B, C, Table~\ref{tab:xmm}). Phase
  intervals indicated as in Fig.~\ref{fig:igrpptot}. The profile of
  \xmm\ observation C required a shift in phase by $-$0.028 for
  alignment with the profile of observation B (consistent with the
  other profiles). The relative normalization is consistent with the
  total pulsed flux (0.8--2.0 keV) being approximately constant (see
  Table~\ref{tab:phaseres}).
\label{fig:softpp1}}
\end{figure}

Fig.~\ref{fig:softpp1} shows \asca\ and \xmm\ pulse profiles in the
energy range 0.8--2.0~keV. This figure strongly suggests that the
pulse profile taken with \asca\ is different from the three
\xmm\ pulse profiles. For \asca, the pulse that peaks around phase 0.6
is higher than the pulse that peaks around phase 1.1, while it is the
other way around for all three \xmm\ profiles. The latter profiles
seem rather similar. We tested this applying the combination of the
Pearson $\chi^2$ and Run tests described in Sect.~\ref{sec:igrpp}. The
\xmm\ profile with the highest statistics (Obs.~B) was used as
template.  \xmm\ profile A, with the lowest statistics, differs from
profile B at the 1.8$\sigma$ level, and profile C differs from profile
B at the 2.6$\sigma$ level. Slight differences in shapes are visible
around the peaks of the two pulses, but we do not consider this
evidence for significant time variability between the
\xmm\ observations.  To the contrary, the \asca\ profile for
0.8--2.0~keV differs from \xmm\ profile B at the 16.6$\sigma$
level. The above-mentioned differences in peak heights in the
\asca\ and \xmm\ profiles become also clear when comparing the fluxes
in these pulses (see Sect.~\ref{sec:softphaseres} and
Table~\ref{tab:phaseres}). The pulse peaking at phase 0.6 contains
$\sim$18\% more flux than the average pulse during the
\xmm\ observations. The pulse peaking at phase 1.1 contains $\sim$34\%
less flux than the average pulse during the \xmm\ observations.

\begin{figure}
\psfig{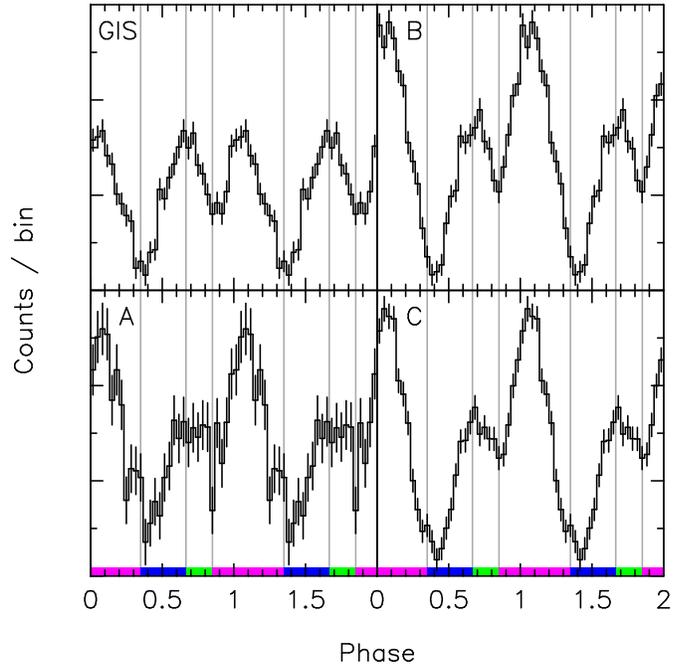}
\caption{\0142\, pulse profiles for 2.0--8.0 keV of \asca\, and the
  three \xmm\, observations (A, B, C, Table~\ref{tab:xmm}). The
  relative normalization scales approximately with the total pulsed
  flux (2--8 keV; see e.g. Table 8). See also the caption of
  Fig.~\ref{fig:softpp1}.
\label{fig:softpp2}}
\end{figure}

Similar conclusions can be reached for the energy band above 2~keV for
which the profiles are shown in Fig.~\ref{fig:softpp2}.
\xmm\ profiles A and C differ from profile B at the 1.3$\sigma$ and
3.5$\sigma$ levels, respectively\footnote{\citet{Rea07_0142xmm},
  analysing the same \xmm\ observations obtained in 2004, also
  concluded that there are no significant differences in the profile
  shapes.}; the \asca\ profile for 2.0--8.0~keV deviates at the
9.3$\sigma$ level. The variation is more drastic than the difference
in profile shape may indicate. The flux in the GIS pulse peaking at
phase 1.1 is 44\% lower than the averaged flux in the same pulse in
the \xmm\ profiles. The flux in the GIS pulse peaking at phase 0.6 is
only 12\% reduced in flux. The difference is evident in
Fig.~\ref{fig:softpp2} in which the profiles are approximately scaled
to the measured 2--8~keV total pulsed flux.

We conclude that we found significant variability in the overall
pulse-profile shapes for energies below 8~keV from the epoch of the
ASCA observations (July/August 1999) to the epochs of the
\xmm\ observations in the years 2003--2004 (see
Table~\ref{tab:xmm}). Particularly in the relative fluxes and spectra
of the two pulses the change in shape is more drastic than reported
before from \pca\ observations by \citet{Dib07_0142rxte} and from more
\xmm\ observations by \citet{Gonzalez07_0142}. \citet{Morii05_0142b}
also report pulse-shape changes from the \asca\ data. They use the
1999 observations (which we also use here) as a template to test two
shorter observations from 1994 and 1998. Due to the low statistical
quality of these data sets only significant changes below 3~keV could
be claimed. \citet{Dib07_0142rxte} notice that the 1998 \asca\ profile
presented by \citet{Morii05_0142b} agrees with the pre-gap RXTE
profile while the 1999 one does not. Interestingly, the profile
measured with {\em Chandra} \citep{Gonzalez07_0142} in May 2000 seems
to be a transition between the \asca\ (1999) and \xmm\ profiles. If
the deviant \asca\ profile is caused by the glitch prior to this
observation as proposed by \citet{Morii05_0142}, then it took about a
year to get to the stable situation as measured with \xmm\ in
2003--2004 and as can be seen from the RXTE pulse profiles
\citep{Dib07_0142rxte}.

Apparently, some relatively fast reconfiguration in
the pulse shape is ongoing during the 1999 \asca\ and 2000 {\em
  Chandra} observations, which might be due to a proposed glitch
\citep{Morii05_0142b} or other drastic event.

\subsubsection{Pulse-profile changes with energy}
\label{sec:ppvar}

\begin{figure}
\psfig{figure=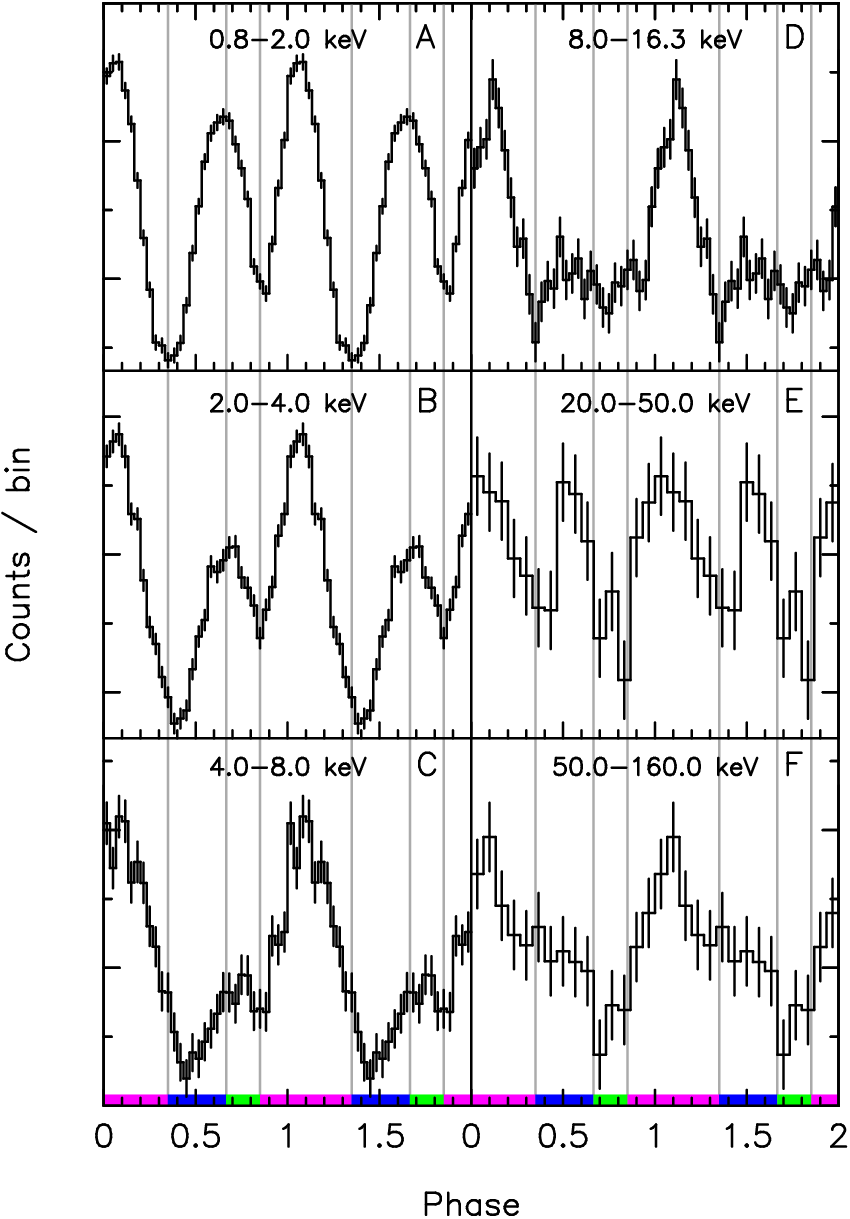,width=\columnwidth,angle=0,clip=t}
\caption{\0142\, pulse profiles from soft to hard X-rays. In panels A,
  B and C, pulse profiles in the energy ranges 0.8--2.0~keV,
  2.0--4.0~keV and 4.0--8.0~keV are shown which are the sums of the
  \xmm\ observations A, B, and C. Panel D presents the \pca\ pulse
  profile in the energy band 8.0--16.3~keV taken from
  \citet{Kuiper06_axps}. Finally, in panels E and F \igr\ pulse
  profiles are shown in the energy ranges 20--50~keV and 50--160~keV.
  Phase intervals are indicated as in Fig.~\ref{fig:igrpptot}.
\label{fig:allpp}}
\end{figure}

\citet{Kuiper06_axps} presented pulse profiles as a function of energy
from \asca, \pca\ and HEXTE. From that work and
e.g. \citet{Israel99_0142,Dib07_0142rxte,Rea07_0142xmm} it is clear
that the pulse profiles significantly change morphology with
energy. In Sect.~\ref{sec:softpp} we have shown that the pulse
profiles of the 2003--2004 \xmm\ observations are statistically the
same. Therefore, we use in this section the summed `time-averaged'
pulse profile using all three \xmm\ observations for comparison with
maximal statistics with the profiles derived at higher energies with
\pca, and \igr.

In Fig.~\ref{fig:allpp} time-averaged pulse profiles from \xmm, \pca,
and ISGRI, are shown in 6 differential energy bands between 0.8~keV
and 160~keV. It is evident that the morphology changes are ongoing
throughout the energy range. One component that is apparent in all
energy bands is the pulse that peaks around phase 1.1. Whereas the
pulse that peaks at phase 0.7 in the \xmm\ band is vanishing and no
longer visible in the \pca\ band, in the \igr\ band a pulse is
visible around phase 0.5. In Sect.~\ref{sec:phaseres} we will
disentangle the contributions from different components using
phase-resolved spectroscopy.


\subsection{Timing analysis: total-pulsed and phase-resolved pulsed spectra}
\label{sec:phaseres}
\begin{figure}
\psfig{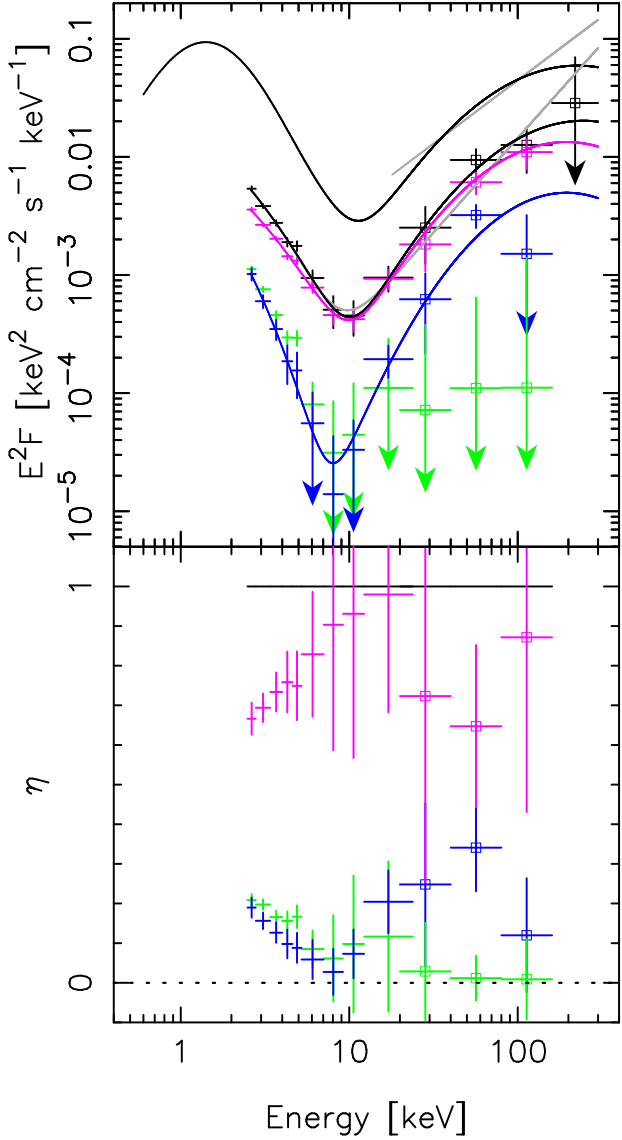}
\caption{In the top panel the phase-resolved pulsed-emission spectra
  and fits of \igr\ and \pca\ are presented. The data points with a
  square symbol are \igr\ measurements and those without a marker are
  from \pca. In black is presented the total-pulsed spectrum; in blue
  Ph\,I; in green Ph\,II; and in magenta Ph\,III. The arrows indicate
  the flux measurements with a significance less than 1.5$\sigma$.
  For comparison the \igr/\xmm-B total-spectrum fit is drawn in
  black. Finally, in grey is plotted the power-law fit for the
  \igr\ total spectrum and the power-law fit for the total-pulsed
  spectrum. In the bottom panel $\eta$ is presented in the same colour
  scheme as the spectra. $\eta$ is defined as the fraction of the
    pulsed emission in a selected phase interval Ph\,I, Ph\,II or
    Ph\,III of the total-pulsed emission, i.e. the sum equals unity.
\label{fig:igrrxte}}
\end{figure}

Total-pulsed spectra were created by the procedure
described in Sect.~\ref{sec:igrtiming}. An example of how the excess
counts in the pulsed components were extracted above a flat background
level is shown in Fig.~\ref{fig:igrpptot}.

We performed phase-resolved spectroscopy for the first time for just
the pulsed emission. \citet{Rea07_0142xmm} also performed
phase-resolved spectroscopy with the \xmm\ data. However, they did not
separate the pulsed component from the dominant (on average
$\sim$85\%) DC component, which has a different spectrum and is
invariant with phase.  For our phase-resolved spectroscopy, we
selected three phase intervals for which we want to extract
high-energy spectra, using the morphology of the ISGRI 20--160 keV
pulse profile (Fig.~\ref{fig:igrpptot}) for defining the phase
boundaries: Ph\,I contains the smaller 2nd pulse in the \igr\, band,
Ph\,II the DC level and Ph\,III contains the main pulse at hard
X-rays, as well as the main pulse for energies below 10~keV. The
definitions of the phase intervals are given in Table~\ref{tab:ph} and
are indicated with grey vertical lines in all figures with pulse
profiles.

\begin{table}[!tbh]
\centering
\renewcommand{\tabcolsep}{1.7mm}
\caption[]{Selected phase intervals for extraction of high-energy
  spectra, using the pulse-shape morphology of the \igr-IBIS-ISGRI
  20--160~keV pulse profile (Fig.~\ref{fig:igrpptot}).}

\begin{tabular}{lll}

\vspace{-3mm}\\

\hline
\hline
\vspace{-3mm}\\
 & Phase interval & Component \\

\hline
\vspace{-3mm}\\
Ph\,I   & $\lbrack0.35, 0.666\rangle$  & Secondary \igr\ pulse \\
Ph\,II  & $\lbrack0.666, 0.851\rangle$ & DC level \igr\ profile \\
Ph\,III & $\lbrack0.0, 0.35\rangle \lor \, [0.851, 1.0\rangle$ &
    Main \igr\ pulse\\

\hline
\end{tabular}

\label{tab:ph}
\end{table}

\subsubsection{INTEGRAL and RXTE}
\label{sec:igrphaseres}
\begin{table}
\centering
\renewcommand{\tabcolsep}{1.0mm}
\caption[]{\pca\ and \igr\ spectral-fit parameters for the
  total-pulsed (TP) spectrum and for three phase intervals (see
  Table~\ref{tab:ph}). The $N_{\rm H}$ is fixed to $0.57 \times
  10^{22}$ cm$^{-2}$. The subscripts below the fit parameters indicate
  if this parameter accounts for the soft (s, $<$10~keV) or hard (h,
  $>$10~keV) X-ray part of the spectrum. The normalizations
  ($F_{0\,{\rm{s}}}$ and $F_{0\,{\rm{h}}}$) are taken at $E_0$ = 1~keV
  and have the units \pcsk.}

\begin{tabular}{lcc}

\vspace{-3mm}\\

\hline
\hline
\vspace{-3mm}\\
TP  & Fit 1 & Fit 2\\

\hline
\vspace{-3mm}\\
$\alpha_{\rm{s}}$  & $2.27 \pm 1.00 $
                  & $2.82 \pm 1.08 $\\
$\beta_{\rm{s}}$   & $1.71 \pm 0.90 $
                  & $1.10 \pm 0.96 $\\ 
$F_{0\,{\rm{s}}}$       & $(1.34 \pm 0.38) \times  10^{-2}$
                  & $(1.79 \pm 1.23) \times  10^{-2}$\\
\vspace{-2mm}\\
$\Gamma_{\rm{h}}$  & $0.40 \pm 0.15 $\\
$\alpha_{\rm{h}}$  & 
                  & $-2.83 \pm 2.16 $\\
$\beta_{\rm{h}}$   & 
                  & $1.00 \pm 0.67 $\\ 
$F_{0\,{\rm{h}}}$       & $(9.17 \pm 4.47) \times  10^{-6}$
                  & $(0.34 \pm 2.62) \times  10^{-8}$\\
\vspace{-2mm}\\
\chir\, (dof) & 0.78 (13)
            & 0.50 (12)\\
$F_{2-10\,\rm{keV}}$ (\ecs) & $(6.82 \pm 1.77) \times 10^{-12}$
                            & $(6.85 \pm 0.93) \times 10^{-12}$ \\
$F_{20-150\,\rm{keV}}$      & $(2.68 \pm 1.34) \times 10^{-11}$
                            & $(2.77 \pm 0.68) \times 10^{-11}$ \\

\hline
\vspace{-3mm}\\
Ph\,I   \\

\hline
\vspace{-3mm}\\
$\alpha_{\rm{s}}$  & $2.90 \pm 2.14 $
                  & $2.89 \pm 3.66 $\\
$\beta_{\rm{s}}$   & $2.56 \pm 2.88 $
                  & $2.42 \pm 3.35 $\\ 
$F_{0\,{\rm{s}}}$       & $(0.67 \pm 1.71) \times  10^{-2}$
                  & $(0.63 \pm 1.74) \times  10^{-2}$\\
\vspace{-2mm}\\
$\Gamma_{\rm{h}}$  & $0.12 \pm 0.20 $\\
$\alpha_{\rm{h}}$  & 
                  & $-4.00 \pm 4.62 $\\
$\beta_{\rm{h}}$   & 
                  & $1.31 \pm 0.72 $\\ 
$F_{0\,{\rm{h}}}$       & $(5.78 \pm 8.81) \times  10^{-7}$
                  & $(0.67 \pm 5.47) \times  10^{-9}$\\
\vspace{-2mm}\\
\chir\, (dof) & 1.43 (13)
              & 0.69 (12)\\
$F_{2-10\,\rm{keV}}$ (\ecs) & $(1.62 \pm 0.60) \times 10^{-12}$
                          & $(1.14 \pm 3.64) \times 10^{-12}$ \\
$F_{20-150\,\rm{keV}}$      & $(5.99 \pm 3.29) \times 10^{-12}$
                          & $(6.89 \pm 6.42) \times 10^{-11}$ \\
\hline
\vspace{-3mm}\\

Ph\,II   \\

\hline
\vspace{-3mm}\\
$\alpha_{\rm{s}}$  & $2.95 \pm 1.71 $\\
$\beta_{\rm{s}}$   & $1.59 \pm 1.56 $\\ 
$F_{0\,{\rm{s}}}$       & $(5.25 \pm 0.72) \times  10^{-3}$\\
\vspace{-2mm}\\
\chir\, (dof) & 0.86 (15)\\
$F_{2-10\,\rm{keV}}$ (\ecs) & $(1.30 \pm 0.03) \times 10^{-12}$ \\

\hline

\vspace{-3mm}\\

Ph\,III   \\

\hline
\vspace{-3mm}\\
$\alpha_{\rm{s}}$  & $1.63 \pm 0.59 $
                  & $2.68 \pm 0.79 $\\
$\beta_{\rm{s}}$   & $2.07 \pm 0.44 $
                  & $1.01 \pm 0.70 $\\ 
$F_{0\,{\rm{s}}}$       & $(5.55 \pm 2.34) \times  10^{-3}$
                  & $(1.02 \pm 0.52) \times  10^{-2}$\\
\vspace{-2mm}\\
$\Gamma_{\rm{h}}$  & $0.50 \pm 0.11 $\\
$\alpha_{\rm{h}}$  & 
                  & $-2.99 \pm 1.73 $\\
$\beta_{\rm{h}}$   & 
                  & $1.09 \pm 0.54 $\\ 
$F_{0\,{\rm{h}}}$       & $(1.11 \pm 0.39) \times  10^{-5}$
                  & $(0.26 \pm 1.67) \times  10^{-7}$\\
\vspace{-2mm}\\
\chir\, (dof) & 1.1 (13)
            & 0.65 (12)\\
$F_{2-10\,\rm{keV}}$ (\ecs) & $(4.72 \pm 0.52) \times 10^{-12}$
                            & $(4.82 \pm 0.47) \times 10^{-12}$ \\
$F_{20-150\,\rm{keV}}$      & $(2.06 \pm 0.82) \times 10^{-11}$
                            & $(2.07 \pm 1.09) \times 10^{-11}$ \\
\hline

\end{tabular}

\label{tab:igrphase}
\end{table}

For the \igr\ data we constructed four broad-band pulse profiles in
the energy ranges 20--40~keV; 40--80~keV; 80--160~keV and
160--300~keV. From these pulse profiles we got three total-pulsed flux
measurements and one upper limit.

For \pca, we extracted 14 pulse profiles in the energy band 2.5--31.5
keV for which we determined the number of excess counts (see
Sect.~\ref{sec:rxte}). The last energy bin (23.8--31.5 keV)
yielded no significant pulse profile, but the \igr\ and \pca\ spectra
nicely bridge the {\it minimum} luminosity.

The time-averaged total-pulsed spectra derived from \igr\ and \pca\
data are presented in Fig.~\ref{fig:igrrxte}. The spectra are fitted
simultaneously using a logparabola for the soft X-rays below 10~keV
and either a power law or a logparabola for the hard X-rays. The fit
results are summarized in Table~\ref{tab:igrphase}. In
Fig.~\ref{fig:igrrxte} the fit with two logparabolae is plotted in
black while the logparabola and a power-law fit is drawn in grey. Both
fits to the \igr\ data are fully acceptable, but the single power-law
model uses less parameters. It can be seen that if the hard X-ray
spectrum has a power-law shape with $\Gamma = 0.4$ that the pulsed
fraction could become as high as 100\% around $\sim$250~keV. If the
spectral shape is logparabolic, the pulsed fraction would not increase
and it would stabilise around $\sim$30\%--40\%

For the three differential phase intervals (see Table~\ref{tab:ph})
also spectra are extracted and plotted in
Fig.~\ref{fig:igrrxte}. Immediately clear is that Ph\,III contains the
bulk of the energy of the pulsed emission. The spectrum shows a
similar transition (from soft to hard) as the total-pulsed
spectrum. Also visible is that the soft X-ray spectrum ($<$10~keV) of
Ph\,III is quite a bit harder than the soft X-ray spectra of Ph\,I and
Ph\,II. While below 10~keV Ph\,I and Ph\,II form one pulse, in the
\igr\, band only in Ph\,I is a pulse visible. From the spectrum and
fit in Fig.~\ref{fig:igrrxte} it can be seen that this transition for
Ph\,I is extremely drastic from the very soft to the very hard.
\begin{figure*}
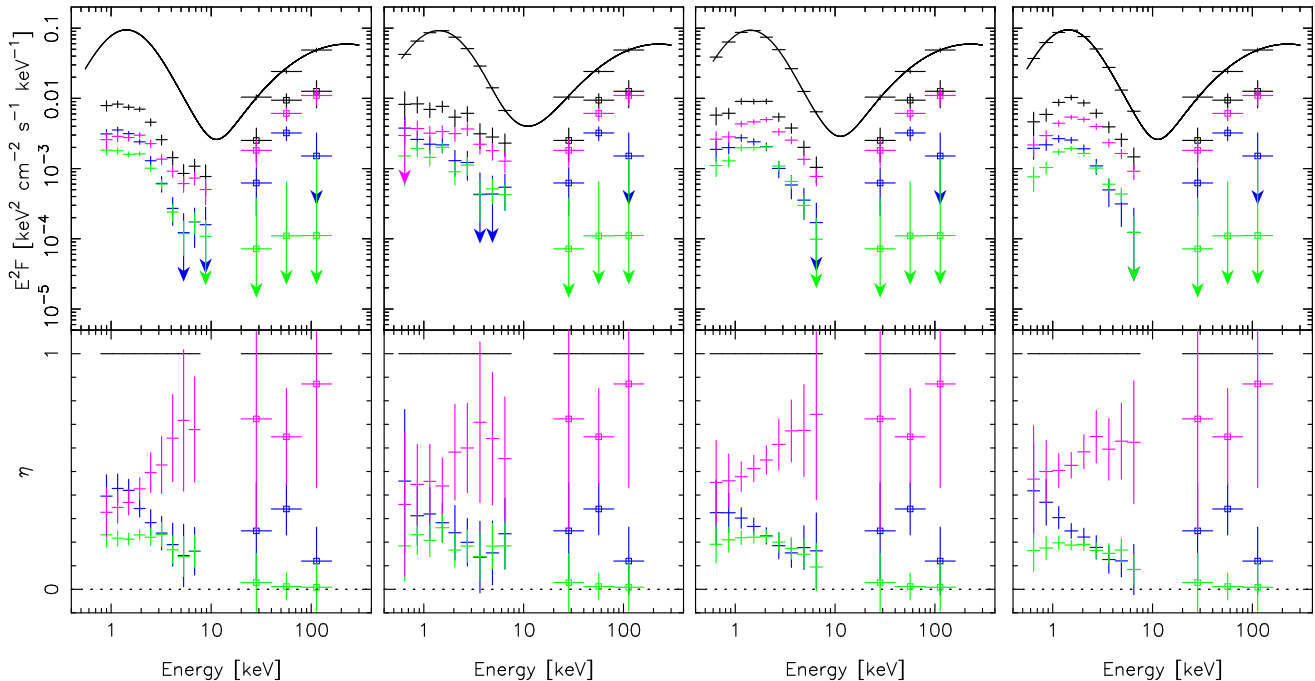

\centering
\begin{minipage}[c]{0.24\textwidth}
\psfig{figure=09390fig09a.ps,height=9cm,angle=0,clip=t}
\end{minipage}
\begin{minipage}[c]{0.24\textwidth}
\psfig{figure=09390fig09b.ps,height=9cm,angle=0,bbllx=105bp,bblly=67bp,bburx=397bp,bbury=700bp,clip=t}
\end{minipage}%
\begin{minipage}[c]{0.24\textwidth}
\psfig{figure=09390fig09c.ps,height=9cm,angle=0,bbllx=105bp,bblly=67bp,bburx=397bp,bbury=700bp,clip=t}
\end{minipage}
\begin{minipage}[c]{0.24\textwidth}
\psfig{figure=09390fig09d.ps,height=9cm,angle=0,bbllx=105bp,bblly=67bp,bburx=397bp,bbury=700bp,clip=t}
\end{minipage}
\caption{For 4 (soft) X-ray and \igr\ observations we present the
  total, total-pulsed and phase-resolved pulsed spectra
  (top panels) and $\eta$ (bottom panels, defined as the fraction of
  the total pulsed emission). The (soft) X-ray data from left to right
  are ASCA GIS, \xmm\ PN obs. A, B and C. All data are equally binned
  for easy comparisons. For the total spectra, the best fit is
  shown. In the top panel for \asca\ the best fit to the total
  spectrum of \xmm\ obs. B and \igr\ are shown for clarity. The total
  spectra and total-pulsed spectra are plotted in black. The phase
  resolved spectra for Ph\,I, Ph\,II and Ph\,III (see
  Table~\ref{tab:ph}) are plotted in blue, green and magenta,
  respectively. The \igr\ phase-resolved spectra are indicated in each
  figure with a square symbol. The data points with arrows indicate
  that the flux values have significances less than 1.5$\sigma$. The
  positive 1$\sigma$ error is drawn. In the bottom panels $\eta$ is
  presented in the same colour scheme as the spectra.}
 \label{fig:eta}
\end{figure*}
%

We can not present an accurate pulsed fraction using \pca\ data. {\it
  The pulsed fraction is defined as the pulsed flux (determined as
  described in Sect.~\ref{sec:obs}) divided by the total flux from the
  point source (pulsed + DC).}  The \pca\ is a non-imaging instrument,
making it difficult to determine a reliable total flux and spectrum
due to confusion with RX~J0146.9+6121, a closeby high-mass X-ray
binary \citep{Motch91_0146}, and underlying Galactic-ridge emission
\citep{Valinia98_ridge}. Alternatively, we present $\eta$ in the
bottom panel of Fig.~\ref{fig:igrrxte}.  {\it $\eta$ is defined as the
  fraction of the pulsed emission in a selected phase interval of the
  total-pulsed emission.}  From the $\eta$ of Ph\,III (magenta) it can
be seen that the contribution of this phase interval to the
total-pulsed emission increases from $\sim$65\% around 3~keV to
practically 100\% around 10~keV. The contributions of Ph\,I and Ph\,II
gradually decrease from $\sim$20\% at 2~keV to (consistent with) no
contribution at 10~keV. The contribution of Ph\,II remains consistent
with 0\% in the \igr\ band (DC level), but the contribution to the
pulsed emission from Ph\,I increases again as a result of the small
2nd peak in the \igr\ band. Because the pulsed contributions in both
phase intervals Ph\,I and Ph\,II similarly decrease to $\sim$0\% at
10~keV, and the pulse shape in the soft X-ray band seems to be of one
component, we infer that the small 2nd pulse in the \igr\ band is
unrelated to the soft X-ray pulse. It likely originates from a
different site (height) in the magnetosphere but appears around the
same phase as the soft X-ray pulse.

It is important to remember that the time-averaged \pca\ spectra are
assembled between March 1996 and September 2003, while the \igr\
observations are taken from December 2003 to August 2007. It is then
remarkable how well the spectra (total pulsed and Ph\,III) connect in
Fig.~\ref{fig:igrrxte}, suggesting stable emission (geometry) above
$\sim$10~keV over 11 years.

\subsubsection{\xmm\ and ASCA}
\label{sec:softphaseres}

\begin{table*}
\centering
\renewcommand{\tabcolsep}{1.7mm}
\caption[]{\asca\, and \xmm\,(A, B \& C) fits for the
  total-pulsed (TP) spectra and for spectra of three phase intervals
  (Ph\,I, II and III, see Table~\ref{tab:ph}) for \0142. The $N_{\rm
    H}$ is fixed to $0.57 \times 10^{22}$ cm$^{-2}$.  For each
  observation the three parameters of the parabola are given. The
  normalizations ($F_0$) are taken at $E_0$ = 1~keV and have the units
  \pcsk. }

\begin{tabular}{lcccc}

\vspace{-3mm}\\

\hline
\hline
\vspace{-3mm}\\
TP  & \asca & A & B & C \\

\hline
\vspace{-3mm}\\
$\alpha$  & $1.88 \pm 0.30 $
          & $2.19 \pm 0.38 $ 
          & $1.24 \pm 0.15 $ 
          & $1.12 \pm 0.17 $\\
$\beta$   & $2.03 \pm 0.48 $
          & $0.65 \pm 0.55 $
          & $2.43 \pm 0.26 $ 
          & $2.41 \pm 0.27 $\\ 
$F_0$     & $(8.49 \pm 0.77) \times  10^{-3}$
          & $(7.93 \pm 0.10) \times  10^{-3}$
          & $(8.17 \pm 0.41) \times  10^{-3}$
          & $(7.98 \pm 0.45) \times  10^{-3}$\\
\vspace{-2mm}\\
\chir\, (dof) & 1.8 (7)
            & 0.29 (6)
            & 0.88 (15) 
            & 0.86 (15)\\
$F_{0.8-2\,\rm{keV}}$ (\ecs)& $(11.47 \pm 0.70) \times 10^{-12}$
                            & $(10.96 \pm 1.38) \times 10^{-12}$
                            & $(12.64 \pm 0.50) \times 10^{-12}$
                            & $(12.72 \pm 0.56) \times 10^{-12}$\\
$F_{2-10\,\rm{keV}}$ (\ecs) & $(5.55 \pm 0.33) \times 10^{-12}$
                            & $(9.02 \pm 1.20) \times 10^{-12}$
                            & $(8.16 \pm 0.35) \times 10^{-12}$ 
                            & $(9.17 \pm 0.39) \times 10^{-12}$ \\

\hline
\vspace{-3mm}\\
Ph\,I   \\

\hline
\vspace{-3mm}\\
$\alpha$  & $1.54 \pm 0.27 $
          & $2.72 \pm 0.47 $ 
          & $1.57 \pm 0.20 $ 
          & $1.59 \pm 0.20 $\\
$\beta$   & $3.76 \pm 0.68 $
          & $0.39 \pm 0.84 $
          & $2.71 \pm 0.45 $ 
          & $2.79 \pm 0.45 $\\ 
$F_0$     & $(3.49 \pm 0.28) \times  10^{-3}$
          & $(2.57 \pm 0.38) \times  10^{-3}$
          & $(2.53 \pm 0.14) \times  10^{-3}$
          & $(2.57 \pm 0.14) \times  10^{-3}$\\
\vspace{-2mm}\\
\chir\, (dof)& 0.66 (7)
            & 0.30 (6)
            & 0.61 (15) 
            & 0.50 (15)\\
$F_{0.8-2\,\rm{keV}}$ (\ecs)& $(4.58 \pm 0.46) \times 10^{-12}$
                            & $(3.29 \pm 0.80) \times 10^{-12}$
                            & $(3.56 \pm 0.27) \times 10^{-12}$
                            & $(3.37 \pm 0.25) \times 10^{-12}$\\
$F_{2-10\,\rm{keV}}$ (\ecs) & $(1.39 \pm 0.14) \times 10^{-12}$
                            & $(1.83 \pm 0.52) \times 10^{-12}$
                            & $(1.50 \pm 0.16) \times 10^{-12}$ 
                            & $(1.36 \pm 0.12) \times 10^{-12}$ \\

\hline
\vspace{-3mm}\\

Ph\,II   \\

\hline
\vspace{-3mm}\\
$\alpha$  & $1.51 \pm 0.31 $
          & $2.24 \pm 0.39 $ 
          & $0.94 \pm 0.18 $ 
          & $0.97 \pm 0.17 $\\
$\beta$   & $2.81 \pm 0.52 $
          & $0.83 \pm 0.63 $
          & $3.27 \pm 0.34 $ 
          & $2.97 \pm 0.31 $\\ 
$F_0$     & $(1.79 \pm 0.16) \times  10^{-3}$
          & $(1.80 \pm 0.23) \times  10^{-3}$
          & $(1.75 \pm 0.10) \times  10^{-3}$
          & $(1.51 \pm 0.08) \times  10^{-3}$\\
\vspace{-2mm}\\
\chir\, (dof) & 1.73 (7)
            & 1.09 (6)
            & 0.96 (15) 
            & 1.35 (15)\\
$F_{0.8-2\,\rm{keV}}$ (\ecs)& $(2.50 \pm 0.26) \times 10^{-12}$
                            & $(2.44 \pm 0.50) \times 10^{-12}$
                            & $(2.77 \pm 0.20) \times 10^{-12}$
                            & $(2.41 \pm 0.18) \times 10^{-12}$\\
$F_{2-10\,\rm{keV}}$ (\ecs) & $(1.13 \pm 0.10) \times 10^{-12}$
                            & $(1.65 \pm 0.39) \times 10^{-12}$
                            & $(1.51 \pm 0.10) \times 10^{-12}$ 
                            & $(1.38 \pm 0.10) \times 10^{-12}$ \\

\hline

\vspace{-3mm}\\

Ph\,III   \\

\hline
\vspace{-3mm}\\
$\alpha$  & $1.87 \pm 0.33 $
          & $1.74 \pm 0.43$ 
          & $1.05 \pm 0.14 $ 
          & $0.95 \pm 0.14 $\\
$\beta$   & $1.36 \pm 0.44 $
          & $0.95 \pm 0.55 $
          & $2.31 \pm 0.21 $ 
          & $2.41 \pm 0.20 $\\ 
$F_0$     & $(3.03 \pm 0.37) \times  10^{-3}$
          & $(3.42 \pm 0.55) \times  10^{-3}$
          & $(3.86 \pm 0.19) \times  10^{-3}$
          & $(3.98 \pm 0.19) \times  10^{-3}$\\
\vspace{-2mm}\\
\chir\, (dof) & 1.72 (7)
            & 0.27 (6)
            & 1.01 (15) 
            & 1.54 (15)\\
$F_{0.8-2\,\rm{keV}}$ (\ecs)& $(4.23 \pm 0.58) \times 10^{-12}$
                            & $(5.09 \pm 1.17) \times 10^{-12}$
                            & $(6.30 \pm 0.37) \times 10^{-12}$
                            & $(6.65 \pm 0.36) \times 10^{-12}$\\
$F_{2-10\,\rm{keV}}$ (\ecs) & $(3.01 \pm 0.28) \times 10^{-12}$
                            & $(5.37 \pm 0.82) \times 10^{-12}$
                            & $(5.13 \pm 0.24) \times 10^{-12}$ 
                            & $(5.63 \pm 0.22) \times 10^{-12}$ \\

\hline

\end{tabular}

\label{tab:phaseres}
\end{table*}

For the three \xmm\ and ASCA observations we also determined the
total-pulsed spectra as well as the pulsed emission
spectra for the three selected phase intervals. Systematic
uncertainties in the response for the timing mode of \xmm\ above
$\sim$8~keV prevented us from using the higher-energy data. For this
reason the observational gap between \xmm\ and \igr\ is about 12~keV.

In the top panels of Fig.~\ref{fig:eta} the results of \asca\ and
\xmm\ obs. A, B and C are presented together with the \igr\ results of
Fig.~\ref{fig:igrrxte}. The colour scheme is the same as in
Fig.~\ref{fig:igrrxte}. In black are drawn the total-pulsed spectra
and (except for \asca) the (binned) total spectra. To the total
spectra of \xmm\ and \igr\ the best fits are drawn. In the panel for
\asca\ the fit to the total spectrum for \xmm\ obs.~C and \igr\ is
drawn.

We have fitted each spectrum with a single logparabolic
function. These fit results are given in Table~\ref{tab:phaseres}.
The parameters of a logparabolic function are not easy to compare, but
the flux values indicate that there are some differences between the
different observations, consistent with our findings in the comparison
of the pulse shapes.

Most remarkable are flux differences between the \asca\, observation
and the \xmm\ ones. \citep[The latter are consistent with the fluxes
reported by][]{Gonzalez07_0142}. While the 0.8--2~keV
 total-pulsed fluxes of all observations are comparable,
the 2--10~keV total-pulsed flux of \asca\ is significantly lower than
the rest. From the phase-resolved spectra it becomes clear that this
difference is due to the significantly lower flux (2--10~keV) in the
main pulse (Ph\,III). In this phase interval the \asca\ 2--10~keV flux
is less than in the \xmm\ observations, while the 2--10~keV flux for
phases Ph\,I and Ph\,II are slightly lower than, or comparable to
those measured with \xmm. We note that the 0.8--2~keV \asca\ flux in
Ph\,III is somewhat lower than in the other observations. Note also
that only for \asca\ the 0.8--2~keV flux in Ph\,I is higher than, or
comparable to, this flux in Ph\,III.

Table~\ref{tab:phaseres} shows that over the three \xmm\ observations
the total-pulsed flux in the 0.8--2~keV and 2--8~keV bands as well as
the fluxes in all three phase intervals are statistically stable.  The
latter is consistent with the conclusion that the shape of the pulse
profiles did not change measurably in this time interval.

\subsubsection{Comparison}
\label{sec:phaserescomp}

To determine the pulsed fraction as a function of energy we have
binned the total spectra in the same binning as the total-pulsed and
phase-resolved pulsed spectra.  The pulsed fraction is defined as the
pulsed flux (determined as described in Sect.~\ref{sec:obs}) divided
by the total flux from the point source (pulsed + DC). For the
instruments \asca\ and \pca\ we do not have accurate total spectra,
therefore we also consider $\eta$.

\begin{figure}
\psfig{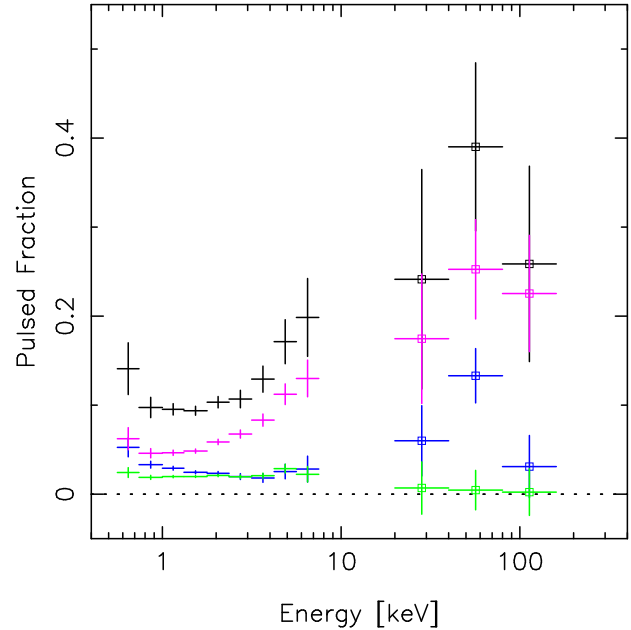}
\caption{Pulsed fraction as function of energy. Below 10~keV the
  averaged \xmm\ pulsed fraction of observations A, B and C are
  plotted. Above 10~keV the \igr\ pulsed fraction is
  plotted. Consistent colour coding as Fig~\ref{fig:igrrxte} and
  Fig.~\ref{fig:eta}; Total-pulsed emission (black), Ph\,I (blue),
  Ph\,II (green), Ph\,III (magenta).
\label{fig:xmmpf}}
\end{figure}

We start with the pulsed fraction. In the previous sections we showed
that \xmm\ observations A, B and C exhibit statistically identical
pulse shapes and pulsed fluxes. Therefore, we have combined these
observations by taking weighted averages.  In Fig.~\ref{fig:xmmpf} the
pulsed fraction as a function of energy is presented for the averaged
\xmm\ and \igr\ observations (black data points). Also presented in
this figure are the pulsed fractions for the three selected phase
intervals (coloured data points). It can be seen in this figure that
the pulsed fraction in the \xmm\ band increases from $\sim$10\% around
1~keV to $\sim$20\% around 8~keV. This trend continues in the \igr\,
band where the pulsed fraction is increasing to possibly as high as
$\sim$40\% in the 40--80 keV band. \citet{Rea07_0142xmm} presented the
pulsed fraction of the fundamental and first harmonic sine functions
for the \xmm\ pulse profiles as a function of energy below 10
keV. These are consistent with the total pulsed fractions we show in
Fig.~\ref{fig:xmmpf}.

Fig.~\ref{fig:xmmpf} also shows that the trend set by the total-pulsed
fraction is followed closely by the pulsed fraction that can be
assigned to Ph\,III (magenta), as we have already discussed in
Sect.~\ref{sec:igrphaseres}. Ph\,III alone accounts for
$\sim$5\%--12\% of the pulsed fraction between 1--8~keV. The
other two phases do not show the same trend. Ph\,I, the 2nd pulse in
the \igr\ band, accounts for $\sim$5\% at 1~keV decreasing to
$\sim$3\% at 8~keV. Finally, Ph\,II is about constant in this energy
range at a level of $\sim$2.5\%. These trends are also visible in the
phase-resolved pulsed spectra in Fig.~\ref{fig:eta}, showing that at
higher energies around 8~keV Ph\,III is dominating over the other two
phase intervals.

\begin{figure}
\psfig{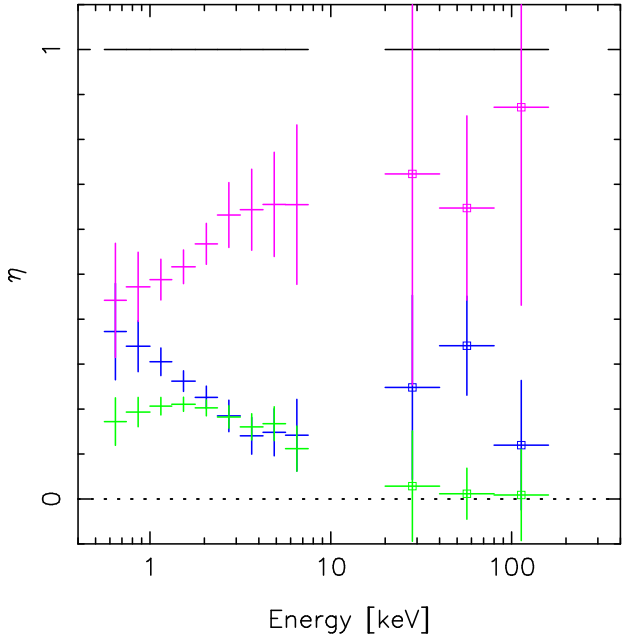}
\caption{Contribution to the total-pulsed emission $\eta$ of
  the three selected phase intervals from the averaged \xmm\ $\eta$
  of observations A, B and C and for \igr\ ($>$10~keV). Colour coding
  consistent with Fig.~\ref{fig:xmmpf}.
\label{fig:xmmeta}}
\end{figure}

To investigate this further, the distribution of $\eta$ is even more
illustrative, as can be seen in Fig.~\ref{fig:eta} and
Fig.~\ref{fig:xmmeta}: the contribution of Ph\,III to the total pulsed
emission is steadily increasing with energy, reaching a fraction of
$\sim$65\% at 8~keV.  For the avarage of the above reported \pca\
observations (Fig.~\ref{fig:igrrxte} the contribution from Ph\,III
increases even to $\sim$100\% around 10~keV.  In the \igr\ band this
contribution is again down to $\sim$60\%--70\%, due to the new pulse
in Ph\,I.

In Fig.~\ref{fig:eta} is also clearly visible that during the \asca\
observation the $\eta$ of Ph\,III below 2 keV is less than that of
Ph\,I. In all observations the $\eta$ of Ph\,I is decreasing with
energy from $\sim$30\%--40\% down to $\lesssim$5\% around
8~keV. However, above 10~keV the contribution of Ph\,I is increasing
again due to the 2nd pulse in the \igr\ band. Ph\,II, which was
chosen as the DC level of unpulsed emission in the \igr\ band, shows
first a constant $\eta$ of $\sim$20\% and it starts decreasing from
2~keV gradually to 0\% in the \igr\ band.

\section{Summary}
\label{sec:sum}
In this paper we have presented new and more detailed characteristics
of \0142 in the hard X-ray regime ($>$10 keV) and studied these in a
broader high-energy view using archival soft X-ray data ($<$10 keV) in
consistent analyses. Many new or more accurate results are reported
which should be considered in theoretical modelling of the high-energy
emission from AXPs.  Particularly new for studies of AXPs is the
performed phase resolved spectroscopy over the total high-energy band,
revealing distinctly different components contributing to the total
emission. We will first present a summary of the results.

\subsection{Total high-energy emission of \0142}

1) ISGRI measures with an effective exposure of 2.4~Ms the most
accurate time-averaged total spectrum between 20 and 230~keV which is
fully consistent with a power-law shape with index 0.93$\pm$0.06 ({\it
  Sect.~\ref{sec:igrtot}, Fig.~\ref{fig:tothigh} and
  Table~\ref{tab:fits}}).

\noindent 2) Combining ISGRI and SPI spectral data provides the first
evidence at the $\sim$4$\sigma$ level for a spectral break with
maximum luminosity at an energy of $\sim$270~keV, assuming a
logparabolic shape of the hard X-ray spectrum ({\it
  Sect.~\ref{sec:igrtot} and Fig.~\ref{fig:tothigh}}).

\noindent 3) The total high-energy spectrum, combining spectra
measured with \xmm\ and ISGRI, can be described satisfactorily with
the sum of three logparabolic functions, two dominating the spectrum
for energies below 8~keV and one above 20~keV \citep[see
  also][]{Rea07_0142}. Minimum values for the luminosity are reached
around 10~keV ({\it Sect.~\ref{sec:xmmtot}, Fig.~\ref{fig:tothigh} and
  Table~\ref{tab:xmmfits}}).

\subsection{Long-term variability in the total emission of \0142}

1) ISGRI does not see evidence for significant long-term variability
of the source flux and power-law spectral index on one-year time
scales between December 2003 and August 2006 at the 17\% (1$\sigma$)
level ({\it Sect.~\ref{sec:igrvar},
  Fig.~\ref{fig:5spec}\,\&\,\ref{fig:contours} and
  Table~\ref{tab:fits}}).

\noindent 2) Two short \igr\ ToO observations following bursting
behaviours of \0142 measured with RXTE show the AXP to be in a state
consistent with the time averaged flux and spectral shape. The total
flux was stable within $\sim$20\% (1$\sigma$; {\it
  Sect.~\ref{sec:igrtoo}}).

\subsection{Pulse profiles}

1) ISGRI measures above 20~keV significant hard-X-ray pulse profiles
up to 160~keV ({\it Sect.~\ref{sec:igrpp} and
  Fig.~\ref{fig:igrpptot}}).

\noindent 2) The morphologies of the profiles measured below 10~keV
with \xmm\ and \asca\ differ from the hard-X-ray profile; one of the
two X-ray pulses vanishes around 10~keV (confirmed with \pca) and a
new pulse appears above 20~keV ({\it Sect.~\ref{sec:ppvar} and
  Fig.~\ref{fig:allpp}}).

\noindent 3) Three \xmm\ observations in 2003--2004 give statistically
identical pulse profiles, but the profile measured by \asca\ in 1999
following a possible glitch of \0142 \citep[see also][]{Morii05_0142b,
  Dib07_0142rxte} is significantly different with different relative
strengths and spectra (0.8--10~keV) of the two pulses visible in the
pulse profiles below 10~keV ({\it Sect.~\ref{sec:softpp} and
  Fig.~\ref{fig:softpp1}\,\&\,\ref{fig:softpp2}}).

\subsection{Pulsed spectra}

1) ISGRI measures above 20~keV a total-pulsed spectrum
up to 160~keV with a shape consistent with a power law with index
0.40$\pm$0.15, significantly different from that of the total
spectrum. This implies that the difference spectrum, the DC component,
is a genuinely different component ({\it Sect.~\ref{sec:igrphaseres},
  Fig.~\ref{fig:igrrxte} and Table~\ref{tab:igrphase}}).

\noindent 2) Below 10~keV, there is evidence for long-term time
variability. The total pulsed fluxes measured for the 0.8--2.0~keV
band with \asca\ in 1999 and in the three \xmm\ observations in
2003--2004 are comparable, but in the 2.0--10.0~keV band different
flux values are measured. The \asca\ value is significantly lower by
$\sim$35\% than the mutually consistent \xmm\ values, and the \pca\,
flux appears to be in between. The $\sim$35\% decrease in total pulsed
flux in the \asca\ observation can largely be assigned to a decrease
in the strength of only one of the X-ray pulses ({\it
  Sect.~\ref{sec:phaseres}, Fig.~\ref{fig:eta} and
  Tables~\ref{tab:igrphase}\,\&\,\ref{tab:phaseres}}). See also
\citet{Dib07_0142rxte} for a significant increased pulsed flux of
$\sim$36\% since 2004 and during the bursting activities in 2006--2007
\citep{Gonzalez07_0142}.

\noindent 3) The total pulsed fraction increases from $\sim$10\% at 1
keV to possibly as high as $\sim$40\% in the 40--80 keV band. Above
these energies this fraction is uncertain, but might become close to
100\%, depending on the actual shapes of the total and pulsed spectra
({\it Sect.~\ref{sec:phaserescomp} and Fig.~\ref{fig:xmmpf}}).

\subsection{Phase-resolved pulsed spectra}

\noindent 1) Phase-resolved spectroscopy over the total high-energy
band reveals the identification of at least three genuinely different
pulse components with different spectra: a) a component in phase
interval Ph\,III (phase 0.85--1.35) peaking at phase 1.1, present and
dominant from 0.8 keV up to 160 keV.  b) a hard-X-ray component in
phase interval Ph\,I peaking in the ISGRI band above 20~keV at phase
$\sim$0.55. c) a soft X-ray component in phase interval Ph\,I + Ph\,II
(phase 0.35--0.85) peaking in the profiles for energies below 10~keV
at phase $\sim$0.65, and not visible above 10~keV.  d) the soft X-ray
component mentioned in c), appears to consist of two parts, a softer
contribution from phase interval Ph\,II than from phase interval
Ph\,I, consistently seen for all reported X-ray observations ({\it
  Sect.~\ref{sec:phaseres}, Fig.~\ref{fig:eta} and
  Tables~\ref{tab:igrphase}\,\&\,\ref{tab:phaseres}}).

\noindent 2) The total pulsed spectra and that of Ph\,III measured
with RXTE (1996--2003) and consecutively with ISGRI (2003--2007)
connect smoothly, and the phase-resolved spectra measured with ASCA
(1999), \xmm\ (2003--2004) and RXTE (1996--2003) are relatively
similar. This is indicative of a remarkably stable geometry of the
contributing emission sites at the surface and/or in the magnetosphere
of the magnetar. There is no evidence for a drastic reordering ({\it
  Sect.~\ref{sec:igrphaseres}, Fig.~\ref{fig:igrrxte} and
  Table~\ref{tab:igrphase}}).


\section{Discussion}
\label{sec:disc}

In this section, we discuss how these new findings fit within the
theoretical models attempting to explain hard X-ray emission from
AXPs. Before the discovery by INTEGRAL of non-thermal emission from
AXPs above 10 keV, \citet{Cheng01_outergap} modeled the production of
high-energy gamma radiation in vacuum gaps proposed to be present in
the outer magnetospheres of AXPs. Gamma-ray emission at the polar caps
will be quenched due to the strong field, but far away from the pulsar
surface gamma radiation could be emitted because the local field will
drop below the critical quantum limit. This radio-pulsar scenario
could offer a stable configuration, suggested by our results to be
required, but their model calculations do not reproduce the hard
spectra and high X-ray luminosities found for e.g. \0142.  After the
discovery by INTEGRAL three approaches were elaborated specifically to
address the non-thermal, very luminous hard X-ray emission from AXPs:
1) a quantum electrodynamics model by \citet{Heyl05_qed,Heyl05_qed2};
2) a corona model by \citet{BT07} and most recently by
\citet{Lyubarsky07_corona}; 3) a resonant upscattering model by
\citet{baring07_london}. Below, we will consider each of these
attempts.

\subsection{Quantum electrodynamics model}
\label{sec:QED}

\citet{Heyl05_qed} created a model to explain the origin of the SGR
and AXP bursts through fast Quantum Electrodynamics (QED) wave modes
which could be triggered when a starquake or other rearrangement in
the magnetic field occurs. In their {\em fast-mode break down} model,
magnetohydrodynamic (MHD) waves generated near the magnetar surface
propagating outwards through the magnetosphere are altered by vacuum
polarization. Hydrodynamic shocks can be developed and
electron-positron pairs produced. An optically thick fireball could be
created emitting mainly X and gamma rays, which can be observed as a
SGR or an AXP burst. To explain stable non-thermal emission
\citet{Heyl05_qed2} extended this model to weaker fast modes that do
not create an optically thick fireball. The resulting spectrum depends
on the total energy delivered by the fast mode and the maximum
magnetic field strength. The latter value determines two energy values
where the spectrum breaks. The peak energy in the spectrum is given by
$E_{{\rm break}} \approx 1600 \frac{B_{{\rm crit}}}{B} mc^2$, where
$B_{{\rm crit}}$ is the quantum critical field of $4.4 \times
10^{13}$\,G, and $B$ the neutron-star magnetic field. The other break
energy is given by $E_0 \approx 10^{-6} \frac{B}{B_{{\rm crit}}}
mc^2$. The spectral shapes below, between and above these break
energies are power-law like with photon indices of  0, 1.5 and 3,
respectively. The power law below $E_0$ can extend down to the optical
regime, in which pulsed non-thermal optical emission
\citep{Eichler02_coherent} has been detected from 4U 0142+61
\citep{Hulleman04_0142, Kern02_0142, Dhillon05_0142}.

Peak energies predicted by this model all exceed 1~MeV. The fast mode
cascade of pairs is not effective if the energy in the pairs ($E_{{\rm
    break}}$) is less than $2mc^2$. This natural limit is called the
{\em minimal model} where $E_{{\rm break}} \approx 1$\,MeV. In this
case, the pairs are created in a (local) magnetic field much weaker
than $B_{{\rm crit}}$.  \citet{Heyl07_london} examined the angular
dependence of the non-thermal emission in the weak field regime
($B_{{\rm NS}} \ll B_{{\rm crit}}$). The produced spectrum strongly
depends on the angle between the magnetic field and the wave
propagation direction.
Specifically, the non-thermal hard X-ray emission is expected to be
largest when the observer's viewing direction to the pulsar is
perpendicular to the local magnetic field, while the thermal emission
is expected to peak when the observer views the polar regions. The
difference between the phases at which the thermal and non-thermal
emissions are peaking can range between 90\degr\ and 180\degr\ if the
angles $\alpha + \beta \ge 90\degr$, where $\alpha$ is the angle
between the rotation axis and the observer's line of sight and $\beta$
the angle between the rotation axis and the magnetic (dipole) axis. If
$\alpha + \beta < 90\degr$, then the phase difference between the
non-thermal and the thermal emission is 180\degr.

Comparing the above QED model predictions with our results, we note
that the power-law spectral index of the total emission above 20 keV
of \0142, 0.93$\pm$0.06, is harder than the predicted value of 1.5.
Furthermore, the estimated break energy of $\sim$270~keV is far below
the allowed minimum value of $\sim$1~MeV, and would mean a neutron
star magnetic field strength of $ 1.3 \times 10^{16}$\,G. The spectral
index of the total pulsed emission of 0.40$\pm$0.15 is inconsistent
with the value 1.5. We note that \citet{Heyl05_qed} do not make a
distinction between the pulsed and total emission.

Finally, there is a clear prediction of a phase difference between the
thermal and non-thermal emission: Our phase-resolved spectroscopy
shows that the actual situation is much more complex: one pulse
exhibits a spectrum composed of a thermal \textsl{and} a non-thermal
component in the same phase interval (Ph\,III); the thermal pulse in
phase interval Ph I+II seems to vanish above 10~keV and a non-thermal
pulse shows up above 20~keV in Ph\,II, also at a negligible phase
difference.

\subsection{Corona model}
\label{sec:corona}

The corona model developed by \citet{BT07} is an extension of earlier
work by \citet{TB05} and \citet{TLK02}.  \citet{TLK02} extended the
magnetar model by a twisted magnetosphere, which was introduced by
\citet{Thompson00_1900} to explain some observational characteristics
seen in SGR~1900+14. Accordingly, the assumed dipolar field gets
twisted due to continuous seismic activity as a result of the extreme
internal magnetic field and a quasi stable loop (tube) can be formed
with a maximum radius of $\sim$2~R$_{{\rm NS}}$. 

Essential in this scenario is an electric field parallel to the
magnetic field which is generated by self induction. This parallel
electric field is capable of accelerating particles and initiates an
electron-positron avalanche. In this way, toroidal field (magnetic)
energy is converted into particle kinetic energy. The energy
production in the tube is related to the voltage difference ($\Phi_e$)
between the footpoints of the loop. The dissipation rate is given by
$L_{{\rm diss}} \sim 10^{37} \Delta\phi(B/10^{15}{\rm G})(a/R_{{\rm
    NS}})(e\Phi_e/{\rm GeV})$ erg\,s$^{-1}$, where $ \Delta\phi$ is
the twist angle (in radians) and $a$ is the size of the twisted
region.  $e\Phi_e$ must be sufficiently high for charged particles to
be lifted from the surface of the neutron star ($\sim$200~MeV) or to
create pairs in the magnetosphere ($\sim$$\gamma_\pm mc^2 = 0.5
\gamma_\pm$~MeV) which are accelerated to $\sim$0.1--1 GeV. 

If the voltage is too low to create a plasma, the tube regulates
itself by either reducing the current or an electric field is
generated ($\partial {\rm {\bf E}}/ \partial t \sim \nabla \times$
{\bf B} -- {\bf j}) until it is high enough to provide the plasma. If
the voltage is too high, the parallel electric field will be screened
by polarization of the plasma. Through this self-regulating voltage
`constant' pair production is achieved, feeding the high-energy
emission. The dissipation time scale of the loop, i.e. the unwinding
of the twist, is of the order 1--10 years. It is determined by the
voltage difference $\Phi_e$ and the corona luminosity $L_{{\rm
    diss}}$. For a global twist of a dipole this becomes: $ t_{{\rm
    decay}} = 0.8 (R_{{\rm NS}} / c^2) L_{{\rm diss}} /
{\Phi_e}^2$. Therefore, the stronger a twist (brighter corona) the
longer the live time.  

The downward beam of electrons and positrons through the tube will
heat a transition layer between the atmosphere and the corona up to
$\sim$200 keV, radiating bremsstrahlung to cool.  In this thermal
transition layer additional pairs can be created after electron ion
collisions.  The emerging spectrum will have a power-law shape with
index $\Gamma \sim 1$ below the break energy. These bremsstrahlung
photons can convert into pairs if the photon energy exceeds $2 m_ec^2/
\sin{\theta}$. In that case synchrotron emission will be
produced. Either way, the maximum energy of photons escaping the
strong magnetic field is $\sim$1~MeV.

\citet{BT07} also address the production of non-thermal emission below
the INTEGRAL-ISGRI energy window. In the 2--10~keV band resonant
upscattering of keV photons can explain the soft power-law like X-ray
tail ($>$5~keV) to the black-body component as first described by
\citet{TLK02} \citep[see also][]{Lyutikov06_compton,Guver06_stems,
  Fernandez07_cyclotron}. For the non-thermal optical and NIR emission
four possible emission mechanisms are discussed with a
curvature-radiation origin being the most plausible explanation. See
also the interpretation of the NIR, optical and soft X-ray
observations by \citet{Hulleman04_0142}, \citet{Morii05_0142} and
\citet{Durant06_0142var}.

A critical aspect in the corona model is the dissipation time scale of
the twist. This determines how long a quasi stable loop can produce a
non-thermal component in the pulse profile. Once the loop vanishes,
the pulse profile should drastically change shape. When a next
starquake produces a new tube, a new non-thermal pulse at a different
phase should appear in the profile.  The unwinding of such quasi
stable loops feeds the non-thermal emission, and it seems plausible
that also long-term flux variability occurs. Our results give
increasing evidence that the non-thermal emission and the global
structure of the pulse profile are stable over many years. During the
INTEGRAL observations covering more than three years, we do not see
significant variation in intensity and spectral shape of the total and
pulsed emission above 20~keV. Furthermore, the RXTE monitoring
observations covering the eight years before the INTEGRAL observations
give time-averaged spectra for the total and total pulsed emission, as
well as for the spectrum of the main pulse (Ph\,III) in the pulse
profile that connect smoothly to the INTEGRAL-ISGRI spectra. In
addition, the pulse maxima and the minima in the pulse profiles
measured by ISGRI and RXTE line-up, suggesting a very stable geometry
and intensity at hard X-rays over more than 10 years. This seems to be
at the limit of what the corona model still can sustain for a bright
corona.

The proposed bremsstrahlung spectrum with index $\Gamma \sim 1$
\citep{BT07} is consistent with our total spectrum for \0142 with
$\Gamma = 0.93\pm 0.06 $ below the break energy.  However, we showed
that the total pulsed spectrum is much harder with $\Gamma =
0.40 \pm 0.15$, and, as a consequence, the DC difference spectrum much
softer, both not consistent with a bremsstrahlung origin. The DC
spectrum extends at least up to 100~keV, given the pulsed fraction of
$\sim 50\% $ at these high energies.

\citet{Lyubarsky07_corona} revisited the heating by the downward beam
of electrons and positrons of the transition region and conclude that
this mechanism is incapable of heating the atmosphere due to
suppression of the required two-stream instability. They propose,
rather, that a hot, tenuous atmosphere forms with a temperature of
$\sim$1--2~MeV, an order of magnitude larger than inferred by
\citet{TB05} and \citet{BT07}. In this hot atmosphere hard radiation
is generated via bremsstrahlung. Furthermore, pairs are easily
produced and fill the whole magnetosphere forming a hot corona.  In
the latter collisionless interaction of the primary beam with the pair
plasma in the corona heats the pairs even more.  Most of this coronal
energy is released in the hard X-ray and soft gamma-ray bands by
Comptonization, so that the overall spectrum of the source is a
superposition of bremsstrahlung radiation from the hot atmosphere and
Comptonization radiation from the extended corona. The latter radiates
in all directions, and the spectrum extends to the MeV band.  The
Comptonization radiation dominates below $\sim$100~keV with photon
spectral slope of $\Gamma \sim$1--2, with a bremsstrahlung spectrum
($\Gamma \sim 1$) above this break energy. This overall spectral shape
is not consistent with our results on \0142. The Comptonization
spectral shape below the break energy might be marginally consistent
with that of the total spectrum, but by far too soft for the total
pulsed spectrum. Furthermore, a bremsstrahlung spectrum after the
break energy would mean an upturn of the spectrum, extending up to the
MeV band. This is inconsistent with the INTEGRAL-SPI results above
100~keV and more importantly, with the COMPTEL upper limits in the MeV
band.  Finally, it is hard to imagine how a Comptonization component
originating in an extended corona radiating in all directions can
produce the phase distribution of \0142 with well defined 
pulses and, for \0142, a sharp minimum.

Our concerns on the long-term stability/variability of the corona
model as expressed for the work by \citet{BT07}, are also valid for
the approach of \citet{Lyubarsky07_corona}.

\subsection{Compton upscattering model}
\label{sec:upscatter}

\citet{baring07_london} consider resonant, magnetic Compton
upscattering to explain the hard X-ray emission. Ultra-relativistic
electrons, accelerated along either open or closed magnetic field
lines, upscatter the thermal soft X-ray atmospheric photons. This
process is very effective close to the magnetar surface
($\lesssim$$10R_{\rm NS}$) where the magnetic field is still strong. For
high-B isolated pulsars it is shown that acceleration of electrons to
ultra-relativistic energies can occur
\citep[e.g.][]{Sturner95_polarcap, Harding98_polarcaps,
  Dyks00_polarcaps}. However, if AXPs with their long periods have
also a dipolar structure at the surface, the acceleration zone cap
radii would become extremely small (of order 50\,m).  As a result, the
relativistic electrons would have to be concentrated in such a narrow
column with charge densities far exceeding the Goldreich-Julian
maximum density \citep{GJ69}. They argue, however, that non-dipolar
structure at the surface is energetically feasible, allowing the
electron acceleration zone to cover a much larger range of
colatitudes than assigned to a standard polar cap.

The Compton scattering formalism invokes full relativistic and
relativistic QED effects in high magnetic fields. Referring to earlier
studies by i.e. \citet{Daugherty86_compton, Ho89_compton} and
\citet{Gonthier00_compton}, and applications in the magnetic Thomson
limit for old gamma-ray burst scenarios \citep{Dermer90_compton,
  Baring94_compton} and for pulsar contexts
\citep{Daugherty89_compton, Sturner95_compton},
\citet{baring07_london} make a first attempt to calculate geometries
and emission spectra produced in the `resonasphere' of AXPs for a
simplified scenario.  They consider the case for a magnetic dipole
field geometry and an injected monoenergetic electron distribution as
well as monoenergetic incident photons of energy commensurate with the
thermal photon temperatures $kT \sim 0.5-1$ keV observed in AXPs.
Important first conclusions are that the Compton resonasphere for long
period AXPs is confined to within few stellar radii of the surface at
higher field locales, and that the emission spectra are considerably
softer than the hard X-ray tails seen in AXPs, and extend to energies
higher than allowed by the COMPTEL upper limits, as e.g. shown in
Fig.~\ref{fig:tothigh}.  The spectra do break above a maximum photon
energy which is dependent on the magnetic field and the Lorentz
factor. Furthermore, the observed spectra depend very sensitively on
the viewing direction of the observer with respect to the
magnetospheric geometry. For investigating such predictions,
phase-resolved spectroscopy as applied in this work offers a very
sensitive diagnostic tool. However, more complicated modelling
addressing non-dipolar field topologies and more realistic electron
and photon spectra are required before our detailed findings can be
confronted with the predictions for hard X-ray emission from resonant
Compton upscattering in AXPs. What can be noted, is that the
preliminary example spectra derived by \citet{baring07_london} are too
soft and extend to energies much higher than can be permitted by the
COMPTEL upper limits.

\subsection{Concluding remark}
\label{sec:remark}

The above discussions indicate that our detailed results as summarized
in the previous section require, for all attempts to model the
non-thermal emission above 20~keV, consideration of a full
three-dimensional geometry taking the relevant angles (magnetic axis
with respect to spin axis, viewing angle) into account, as well as the
physical production processes taking place on the surface of the
neutron star and in different sites of the atmosphere and
magnetosphere. Such analysis would be similar to what is done to model
the non-thermal emission from radio pulsars. Our results seem to
indicate that the non-thermal persistent emission from AXPs is
persistent, requiring a scenario with a stable geometry. To obtain
further information on the geometry, including more information for
the different relevant angles, similar detailed results for the other
non-thermally emitting AXPs \citep[see][]{Kuiper06_axps} are important
to give further constraints on the theoretical modelling.

\begin{acknowledgements}
We would like to acknowledge: C.~Winkler, INTEGRAL Project Scientist
and the INTEGRAL team for the quick ToO response; F.~Haberl for useful
discussions on how to analyse the XMM-Newton timing data
correctly. This work is supported by NWO, Netherlands Organisation for
Scientific Research. RD is supported by the Natural Sciences and
Engineering Research Council (NSERC) PGSD scholarship. FPG is
supported by the NASA Postdoctoral Program administered by Oak Ridge
Associated Universities at NASA Goddard Space-Flight
Center. Additional support was provided by NSERC Discovery Grant Rgpin
228738-03, FQRNT, CIFAR, and
CFI. VMK holds a Lorne Trottier Chair in Astrophysics \& Cosmology, a
Canada Research Chair and is a R.~Howard Webster Foundation Fellow of
CIFAR.  The results are based on observations with INTEGRAL, an ESA
project with instruments and science data centre funded by ESA member
states (especially the PI countries: Denmark, France, Germany, Italy,
Switzerland, Spain), Czech Republic and Poland, and with the
participation of Russia and the USA. The SPI project has been
completed under the responsibility and leadership of CNES.  We are
grateful to ASI, CEA, CNES, DLR, ESA, INTA, NASA and OSTC for
support. This research has made use of data obtained through the
High-Energy Astrophysics Center Online Service, provided by the
NASA/Goddard Space-Flight Center.
\end{acknowledgements}

\bibliography{../literature}
\end{document}